\pdfoutput=1

\documentclass[11pt]{article}

\usepackage[final]{acl}

\usepackage{times}
\usepackage{latexsym}

\usepackage[T1]{fontenc}

\usepackage[utf8]{inputenc}

\usepackage{microtype}

\usepackage{inconsolata}

\usepackage{graphicx}
\usepackage{graphbox}
\usepackage{adjustbox}

\usepackage{subfig}
\usepackage{caption}
\usepackage{subcaption}
\usepackage{amsmath} 
\usepackage{booktabs}
\usepackage{multicol}
\usepackage{multirow}
\usepackage{ragged2e}
\usepackage{enumitem}
\usepackage{array}
\usepackage[titletoc]{appendix}
\usepackage[hang,flushmargin]{footmisc}
\usepackage{cleveref}

\usepackage[frozencache=true,cachedir=minted-cache]{minted} 
\definecolor{LightGray}{gray}{0.9}
\PassOptionsToPackage{dvipsnames}{xcolor}

\title{UniRAG: Universal Retrieval Augmentation \\for Large Vision Language Models}

\author{Sahel Sharifymoghaddam\thanks{~~~Equal contribution.}, Shivani Upadhyay$^{*}$, Wenhu Chen, Jimmy Lin \\[1ex]
       Department of Computer Science\\
   University of Waterloo\\[1ex]
 \texttt{\{sahel.sharifymoghaddam, sjupadhyay, wenhu.chen, jimmylin\}@uwaterloo.ca}}

\begin{document}
\maketitle
\begin{abstract}
Recently, Large Vision Language Models (LVLMs) have unlocked many complex use cases that require Multi-Modal (MM) understanding (e.g., image captioning or visual question answering) and MM generation (e.g., text-guided image generation or editing) capabilities.
To further improve the output fidelity of LVLMs we introduce UniRAG, a plug-and-play technique that adds relevant retrieved information to prompts as few-shot examples during inference.
Unlike the common belief that Retrieval Augmentation (RA) mainly improves generation or understanding of uncommon entities, our evaluation results on the MSCOCO dataset with common entities show that both proprietary models like GPT-4o and Gemini-Pro and smaller open-source models like LLaVA, LaVIT, and Emu2 significantly enhance their generation quality when their input prompts are augmented with relevant information retrieved by Vision-Language (VL) retrievers like UniIR models. 
All the necessary code to reproduce our results is available at \url{https://github.com/castorini/UniRAG}
\end{abstract}

\begin{figure*}[tbh]
  \centering
   \includegraphics[width=\linewidth]{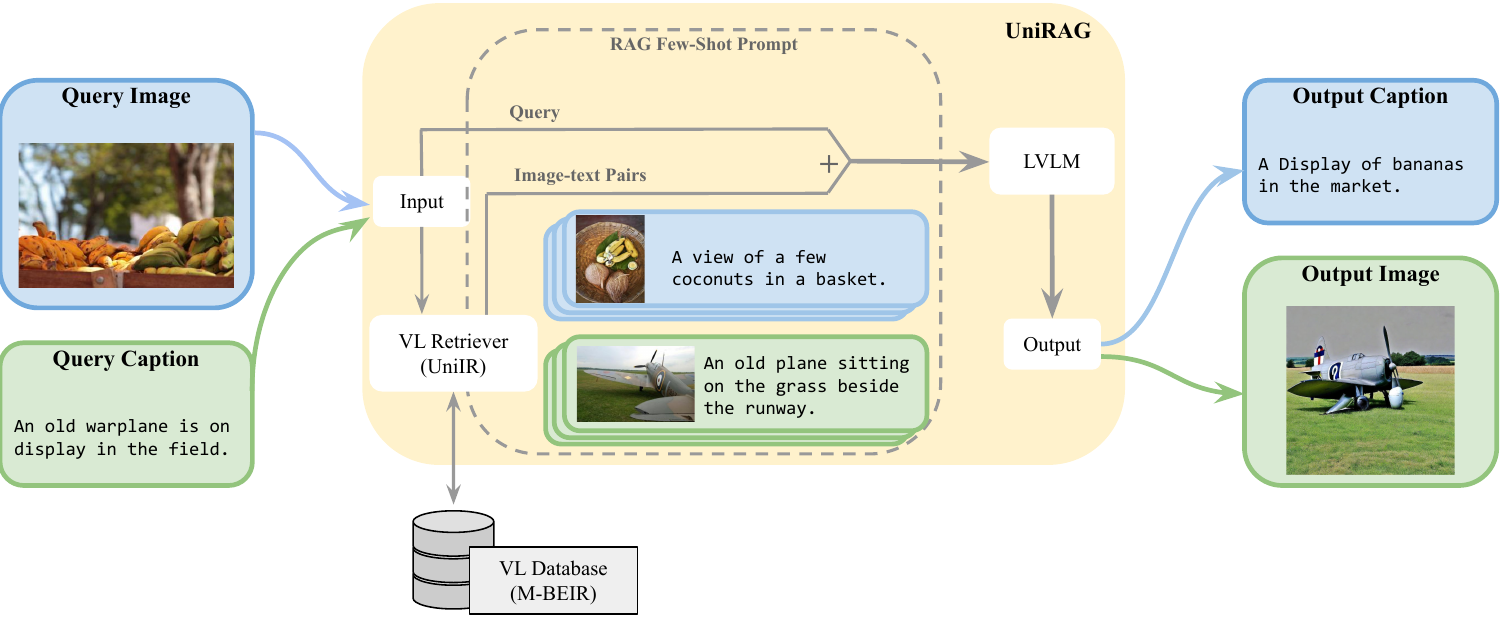}
   \caption{An Overview of the UniRAG technique with image captioning (blue) and image generation (green) tasks. UniRAG retrieves relevant image-text pairs and adds them as few-shot examples to the LVLM's input prompt.}
   \label{fig:flowchart}
\vspace{0.4cm}
\end{figure*}

\section{Introduction}
Recent advancements in Large Vision Language Models (LVLMs), encompassing both proprietary models like GPT-4o~\cite{openai2024gpt4}, Gemini-Pro~\cite{geminiteam2024gemini}, DALL-E~\cite{dalle} and Parti~\cite{parti} and open-source models such as LLaVA~\cite{liu2023visual, liu2023improved}, LaVIT~\cite{lavit} and Emu2~\cite{emu2}, are instrumental in bridging the gap between different modalities.
These models have demonstrated remarkable human-like efficacy and achieved state-of-the-art effectiveness in various benchmark assessments~\cite{mm-benchmark-survey}.

However, like LLMs, VLMs often struggle to produce accurate results on lesser-known or recent topics~\cite{visualhallucinations, mmhallucinations}. 
This limitation occurs because models generate responses solely based on their training data, lacking access to external information during inference. Given the rapidly changing nature of the world, it is unrealistic for models to correctly address every query using only pre-trained knowledge. 

As a workaround, techniques like Retrieval Augmented Generation (RAG) have gained significant attention in recent years~\cite{RAGSurvey}.
In-context RAG~\cite{in-context-rag} particularly relies on the learning capabilities of models to incorporate the latest knowledge and guide them in generating more relevant results during inference. Recent studies have applied RAG to VL tasks, including Visual Question Answering (VQA), text-guided image generation and editing, and image captioning~\cite{ra-cm3, ok-vqa, murag, CM3Leon, reveal}.
However, these approaches mainly focus on training models with large image-text datasets, leaving the potential for plug-and-play in-context RAG in VL applications largely unexplored.
As black-box LVLMs become more prevalent, addressing this gap is essential for their reliable adoption in applications that cannot tolerate hallucinations.

This paper introduces the plug-and-play UniRAG technique, designed to enhance the fidelity of model outputs.
UniRAG integrates RA with LVLMs to guide the generation process using relevant in-context information.
By incorporating interleaved image-text pairs as few-shot examples during inference, UniRAG serves as a model-agnostic approach that effectively teaches out-of-the-box LVLMs about various task intents and uncommon entities.
Our evaluation in this work primarily focuses on image captioning (image-to-text) and image generation (text-to-image) tasks.

UniRAG adopts a two-stage retrieval and generation approach. For retrieval we use the UniIR~\cite{uniIR} retriever, which is trained with diverse VL datasets to retrieve heterogeneous outputs in both text and image modalities.
We mainly adopt UniIR's CLIP Score Fusion and BLIP Feature Fusion models which are instruction-tuned using CLIP~\cite{clip} and BLIP-2~\cite{blip}, respectively.
In the generation stage, we use a variety of off-the-shelf proprietary and open-source LVLM models. 
In particular, we employ LLaVA, GPT-4o and Gemini-Pro for image captioning and LaVIT and Emu2 for image generation.
We guide these LVLMs using zero-shot and few-shot prompting techniques to demonstrate their enhanced generation capabilities due to access to relevant examples during inference.

For our evaluations, we mainly utilize the MSCOCO~\cite{mscoco} dataset from the M-BEIR~\cite{uniIR} benchmark. We assess caption and image generation using M-BEIR's image-to-text and text-to-image tasks, respectively.
By combining RA with LVLMs, we achieve significant improvements over baseline effectiveness: an average increase of about ~9 percentage points in SPICE~\cite{anderson2016spice} for image captioning and a reduction of 25 units in Fréchet Inception Distance (FID)~\cite{fid} for image generation. 
Additional evaluations using M-BEIR's Fashion200k dataset demonstrate that UniRAG is even more effective in domain-specific applications. To summarize, our main contributions in this paper are as follows:
\vspace{-0.1cm}
\begin{itemize}[leftmargin=*]
    \item Introducing the plug-and-play UniRAG technique, which combines VL retrievers with LVLMs using in-context RAG. 
    \item Assessing the effectiveness of UniRAG in image captioning and generation tasks using five LVLMs, including both open-source and proprietary models.
    \item Evaluating UniRAG on the Fashion200k dataset to showcase its effectiveness in domain-specific applications.
\end{itemize}

\section{Related Work}

\paragraph{Retrieval Augmentation with Generative Models:}
Retrieval Augmentation (RAG) techniques have significantly improved the effectiveness of LLM generation by incorporating external information, leading to higher-quality results~\cite{RAGSurvey}. While some researchers modify LLM architectures or fine-tune them for conditioning on retrieved information~\cite{realm, retro}, others explore in-context RAG~\cite{in-context-rag, ewek-qa}, which incorporates relevant data directly into prompts. Recent studies have also examined the integration of RAG with chain-of-thought~\cite{cot, selfconsistency} and other prompting techniques~\cite{knowledge-cot, chainofknowledge}; however, most of these works mainly augment generation with text.

\paragraph{Vision-Language Models and Retrieval Augmentation:}
Recently, a number of models have been developed that integrate vision and language understanding~\cite{dalle,clip, parti, blip, liu2023visual, liu2023improved,lavit,emu2, siglip, openai2024gpt4, geminiteam2024gemini}. These models typically use modality-specific encoders trained on large image-text datasets to form a unified feature space. More details on the LVLMs selected for evaluation can be found in~\Cref{sec:generator}.

Similar to traditional LLMs, LVLMs also struggle with generating content about lesser-known entities or uncommon combinations of common ones~\cite{reimagen, ra-diffusion}. By leveraging retrieval, LVLMs can be guided to produce more accurate and high-fidelity outputs. UniIR~\cite{uniIR} introduces VL retriever models that are instruction-tuned with diverse VL datasets to perform heterogeneous retrieval tasks involving both text and images.
RA-CM3~\cite{ra-cm3}, MuRAG~\cite{murag},  CM3Leon~\cite{CM3Leon} and REVEAL~\cite{reveal} employ a retriever-generator workflow for RAG in various VL tasks.
However, they all require training generator and/or retriever models on extensive VL datasets.

In contrast, UniRAG incorporates a VL retriever during inference, facilitating in-context RAG for black-box LLMs.
The work by~\citet{rag-mm-cot} is the closest to ours, introducing an RA Chain-of-Thought (CoT) method.
However, unlike UniRAG, it focuses on enhancing LVLMs for VQA.
Their approach selects the most relevant task from each VQA benchmark’s training/test set as a few-shot CoT demonstration.
By comparison, UniRAG does not rely on task-specific examples to guide the generator.
Instead, it retrieves relevant information from an external database to enrich the model’s knowledge during inference. This task-agnostic design allows UniRAG to benefit various image-text tasks, including VQA, image captioning, image generation, and Optical Character Recognition (OCR), by incorporating semantically relevant image-text pairs as few-shot examples.




\section{Methodology}
UniRAG adopts a two-stage retrieval and generation workflow that is explained below:

\subsection{Retrieval}
The retrieval stage is where for a given query its top $k$ most relevant candidates are extracted from a VL database. 
In this stage query and candidate modalities can be any combination of $\{text, image\} \rightarrow \{text, image\}$.
Specifically, for an image-to-text task such as caption generation, the query is an image, and retrieved candidates are captions; while for a text-to-image task (e.g., image generation), the query is a caption and retrieved candidates are images.

To use image-text pairs as in-context examples in the next stage, we need to convert the retrieved single-modality candidates into pairs. This involves treating each retrieved candidate as a query to find its complementary candidate based on two criteria: a) it must have the opposite modality (text for an image candidate and vice versa) and b) it must differ from the original query to prevent revealing the ground-truth answer. The retriever models used for our experiments are further explained in~\Cref{sec:setup}.

\subsection{Generation}
\label{sec:generator}
The generation stage is where the LVLM generates the required output in zero-shot~\cite{kojima2023large} or few-shot~\cite{brown2020language} settings. Similar to the retrieval stage, in the zero-shot setting (baseline 1), the query is only present in the input prompt; while in the few-shot setting, in-context examples are additionally included in the prompt. In this setting, examples are obtained using one of the following methods: 1) random selection of queries and their ground-truth candidates from the dataset (baseline 2); 2) in-context RAG with retrieved candidate pairs from the previous stage.
We leverage various LVLMs for caption and image generation tasks, as detailed in~\Cref{sec:setup}. 

\section{Experimental Setup}
\label{sec:setup}
\begin{table*}[thb]
\centering
\resizebox{\textwidth}{!}{%
\begin{tabular}{c | p{4.5cm} | p{5cm} p{5cm} p{5cm}}
\toprule
&
\centering \textbf{Image Query} &
\multicolumn{3}{c}{\textbf{Top retrieved image-caption pair $(k=1)$}}\\
&
\centering Zero-shot &
\centering CLIP-SF &
\centering BLIP-FF &
\centering\arraybackslash Random\\
\midrule
\multirow{4}{*}{{Prompt}} &
\centering \includegraphics[align=c,width=0.2\textwidth]{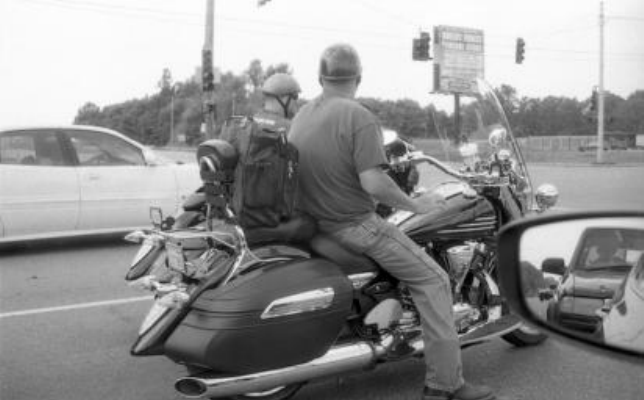}
&
\centering \includegraphics[align=c,width=0.2\textwidth]{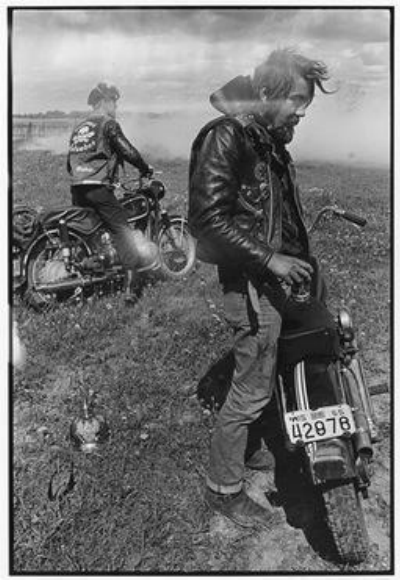}
&
\centering \includegraphics[align=c,width=0.2\textwidth]{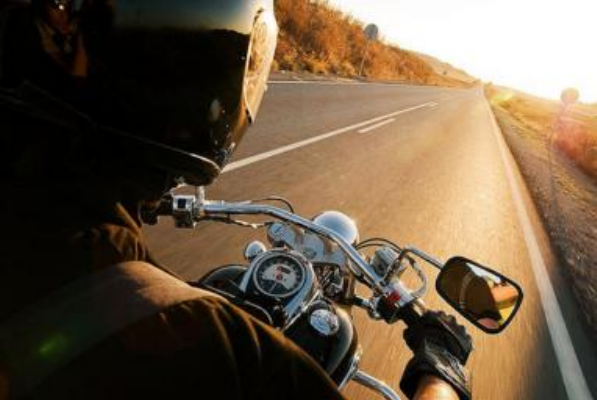}
&
\centering\arraybackslash \includegraphics[align=c,width=0.2\textwidth]{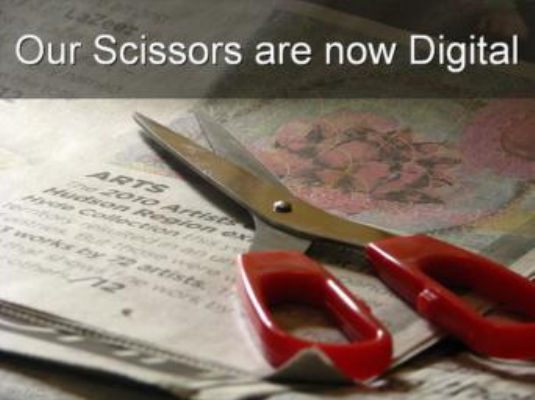}
\\
\cmidrule{2-5}
&
\centering \multirow{3}{*}{{-}}&
Black and white photograph of two men on motorcycles. &
A man riding on the back of a motorcycle down a highway. &
A pair of red scissors sitting on a newspaper. \\
\midrule
\midrule
\multirow{5}{*}{{LLaVA}} &
Motorcycle rider on the road, with luggage attached to the back.&
Black and white photograph of a man on a motorcycle.&
A man riding on the back of a motorcycle down a highway.&
A man is sitting on a motorcycle, and there are two and a half hotdogs on paper plates on a counter.\\
\midrule
\multirow{4}{*}{{Gemini-P}} &
Biker and passenger riding down the road together.&
Black and white photograph of a man and woman riding on a motorcycle.&
Two people riding on a motorcycle through a parking lot.&
Biker and his passenger riding down the road.\\
\midrule
\multirow{4}{*}{{GPT-4o}} &
two people on a motorcycle waiting at a traffic light.&
black and white photograph of two men on a motorcycle at an intersection.&
two people riding on a motorcycle through an intersection.&
two people are sharing a large cruiser motorcycle at a stoplight.\\
\bottomrule
\end{tabular}%
}
\caption{Sample caption generation with LLaVA, Gemini-Pro, and GPT-4o models in zero-shot and one-shot settings. The ``Prompt'' row shows the zero-shot image query as well as retrieved image-caption pairs from CLIP-SF, BLIP-FF and random selection that are included in the prompt as in-context examples.}
\vspace{0.2cm}
\label{tab:caption-retriever}
\end{table*}

\subsection{Selected Models}
We use UniIR's CLIP Score Fusion (CLIP-SF) and BLIP Feature Fusion (BLIP-FF) models as VL retrievers. Experimental results from the original paper indicate that these instruction-tuned UniIR models significantly outperform their pre-trained CLIP and BLIP2 baselines across various tasks and configurations, including image and caption retrieval tasks~\cite{uniIR}.

To generate appropriate text captions for images, we use the LLaVA (13B)~\cite{liu2023improved}, Gemini-Pro ~\cite{geminiteam2024gemini}, and GPT-4o~\cite{openai2024gpt4} models.
\Cref{tab:caption-retriever} presents sample captions generated by these models in zero-shot, random few-shot, and RAG few-shot settings (retrieved by CLIP-SF and BLIP-FF). 

\textit{LLaVA} is an open-source LVLM based on Vicuna~\cite{vicuna}, utilizing CLIP as its visual encoder. It is fine-tuned on VL instruction-following data generated by GPT-4~\cite{liu2023visual}.
LLaVA's ability to understand and follow human intent makes it well-suited for RAG caption generation.  
\Cref{fig:zero-shot-prompt,fig:few-shot-llava-prompt,fig:few-shot-gemini-prompt,fig:few-shot-gpt-prompt} in Appendix~\ref{sec:prompt-details} show detailed prompts used for caption generation.

Unlike image captioning, which only requires MM understanding, image generation necessitates models with both MM understanding and generation capabilities. To meet this need, we selected LaVIT and Emu2-Gen from a survey of LVLMs~\cite{mmllmSurvay}.

\textit{LaVIT} is an open-source model based on LLaMA-7B~\cite{llama} that employs a visual tokenizer to convert images into discrete visual tokens. This allows it to process both text and image inputs using a unified objective for next image/text token prediction. The authors report a Frechet Inception Distance (FID) of 7.4~\cite{fid} on the MSCOCO dataset, indicating its strong capability in image generation tasks.

\textit{Emu2-Gen} is another open-source model, fine-tuned from Emu2 for image generation. Like LaVIT, it uses next token prediction across all modalities and incorporates a variant of CLIP as its visual encoder along with LLaMA-33B as its backbone. It can accept interleaved inputs of images, texts, and videos, and features a visual decoder for generating images. While its zero-shot effectiveness on various VL tasks is lower than that of Emu~\cite{emu}, Flamingo~\cite{flamingo}, and others, it excels in few-shot settings, making it suitable for image generation with RAG.

\Cref{tab:image-retriever} displays visual examples of image generation using LaVIT and Emu2-Gen in zero-shot, random few-shot, and RAG few-shot settings (using UniIR retrievers).
\begin{table*}[thb]
\centering
\resizebox{\textwidth}{!}{%
\begin{tabular}{c | p{4.2cm} | p{4.2cm} p{4.3cm} p{4.5cm}}
\toprule
&
\centering \textbf{Caption Query} &
\multicolumn{3}{c}{\textbf{Top retrieved image-caption pair $(k=1)$}}\\
&
\centering Zero-shot &
\centering CLIP-SF &
\centering BLIP-FF &
\centering\arraybackslash Random\\
\midrule
\multirow{4}{*}{{Prompt}} &
\centering -&
\centering \includegraphics[align=c,width=0.18\textwidth]{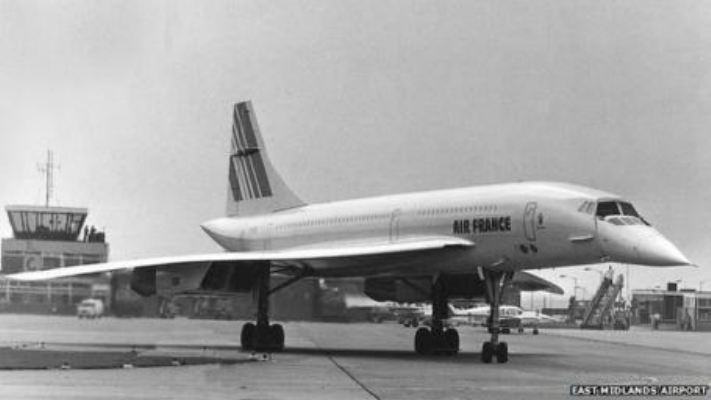}
&
\centering \includegraphics[align=c,width=0.18\textwidth]{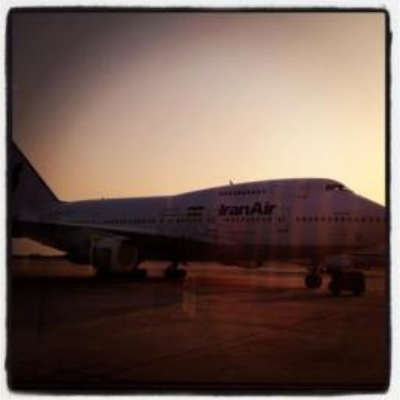}
&
\centering\arraybackslash \includegraphics[align=c,width=0.18\textwidth]{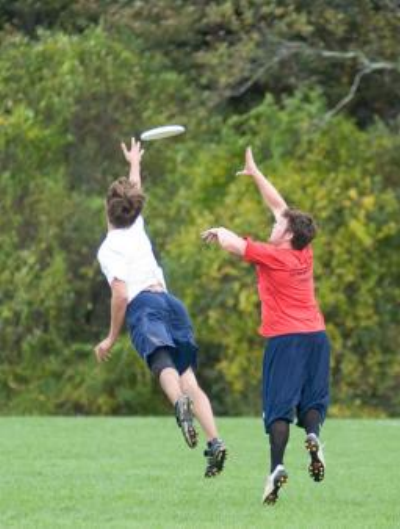}
\\
\cmidrule{2-5}
&
A large jetliner sitting on top of an airport runway.&
Black and white jet airliner on an airport runway.&
A large plane is parked on the run way.&
Two teenage boys are playing a game of frisbee together.\\
\midrule
\midrule
LaVIT &
\centering \includegraphics[align=c,width=0.18\textwidth]{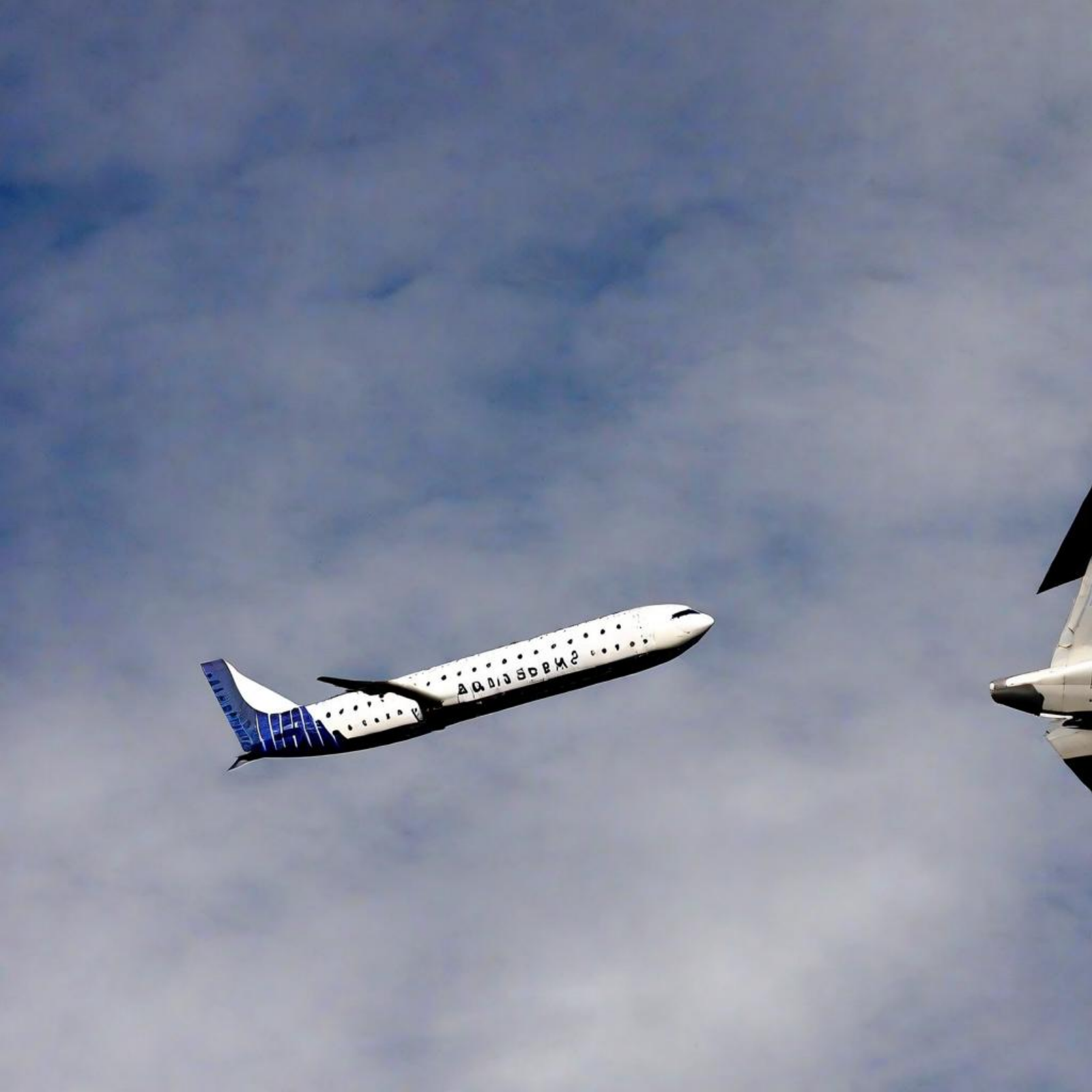}&
\centering \includegraphics[align=c,width=0.18\textwidth]{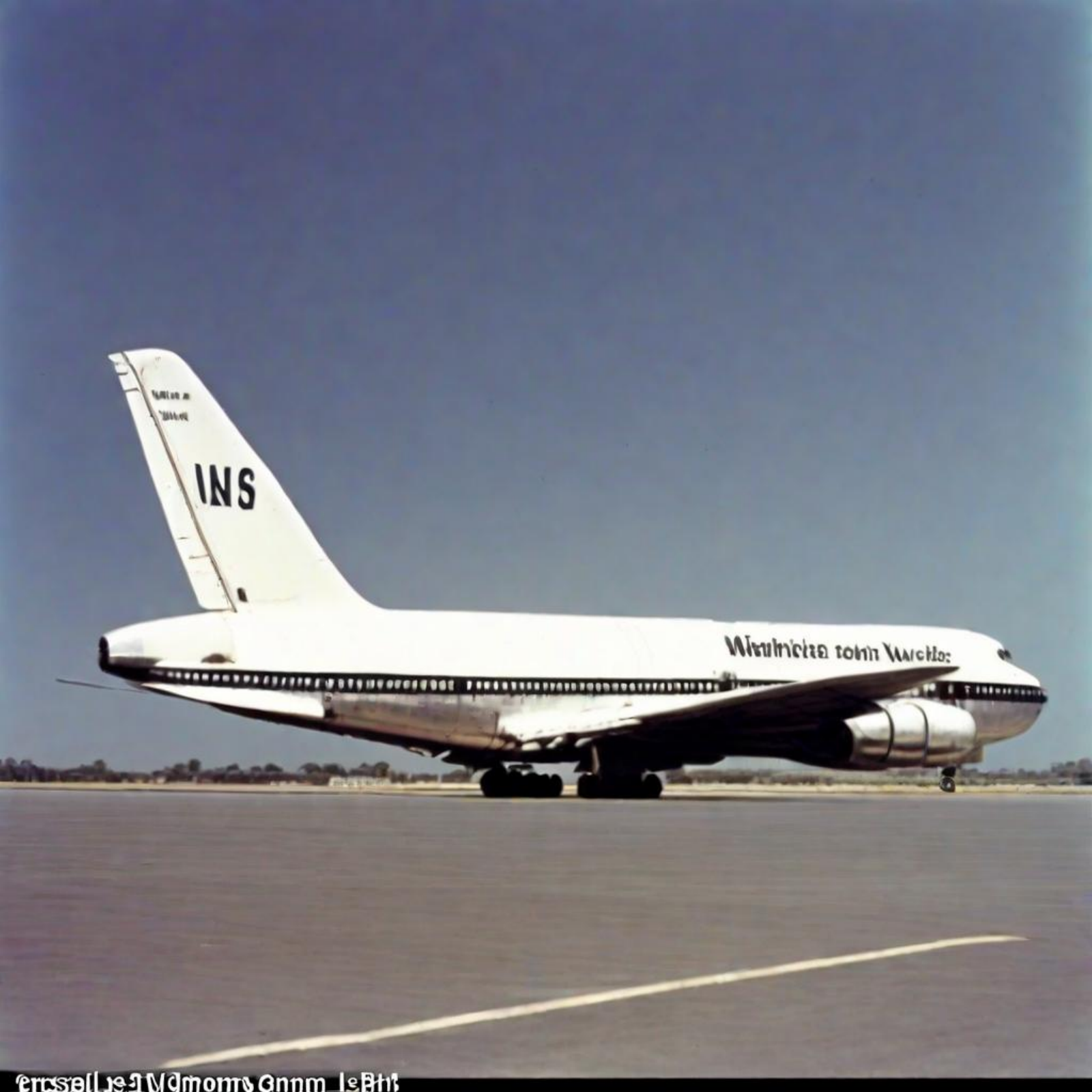}&
\centering \includegraphics[align=c,width=0.18\textwidth]{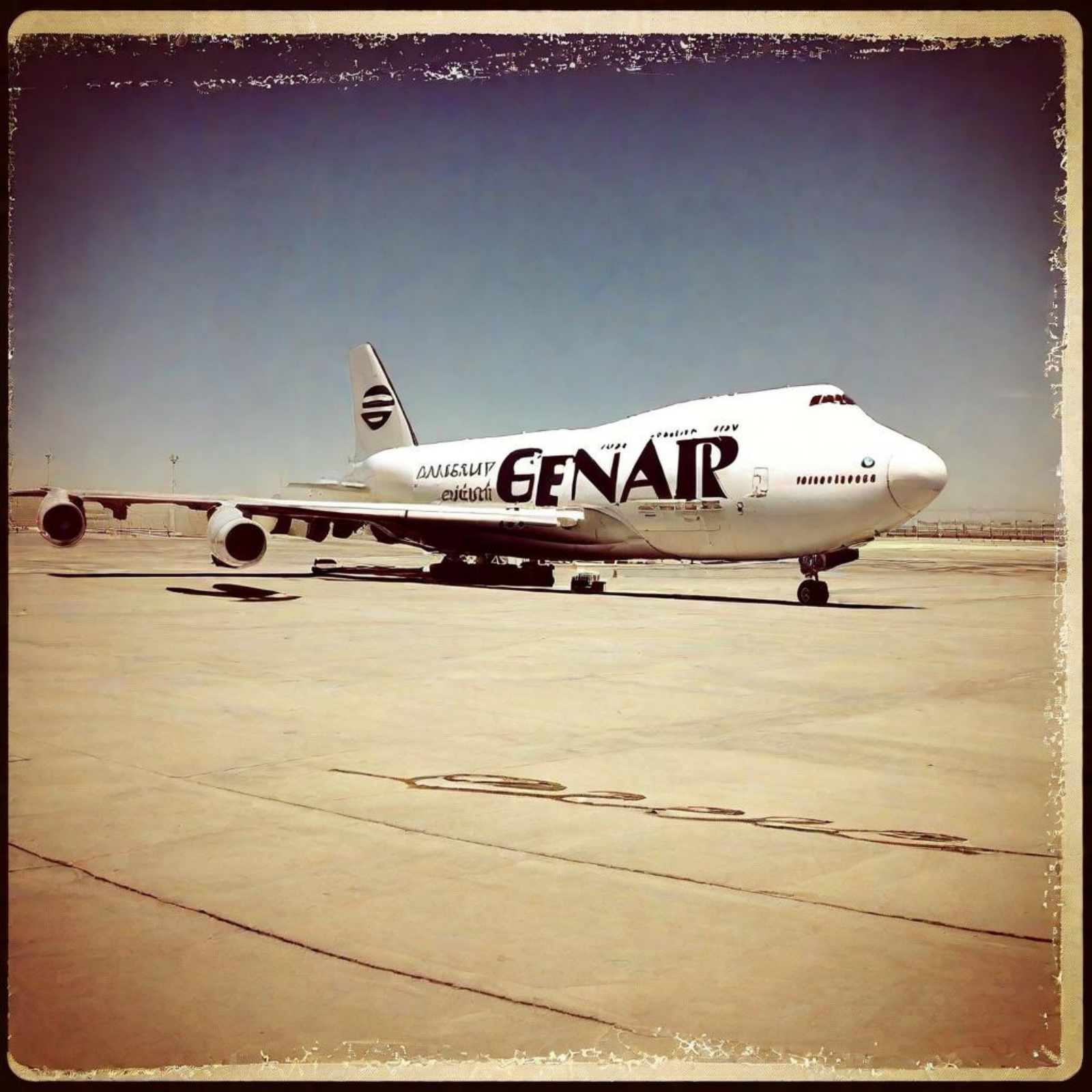}&
\centering\arraybackslash \includegraphics[align=c,width=0.18\textwidth]{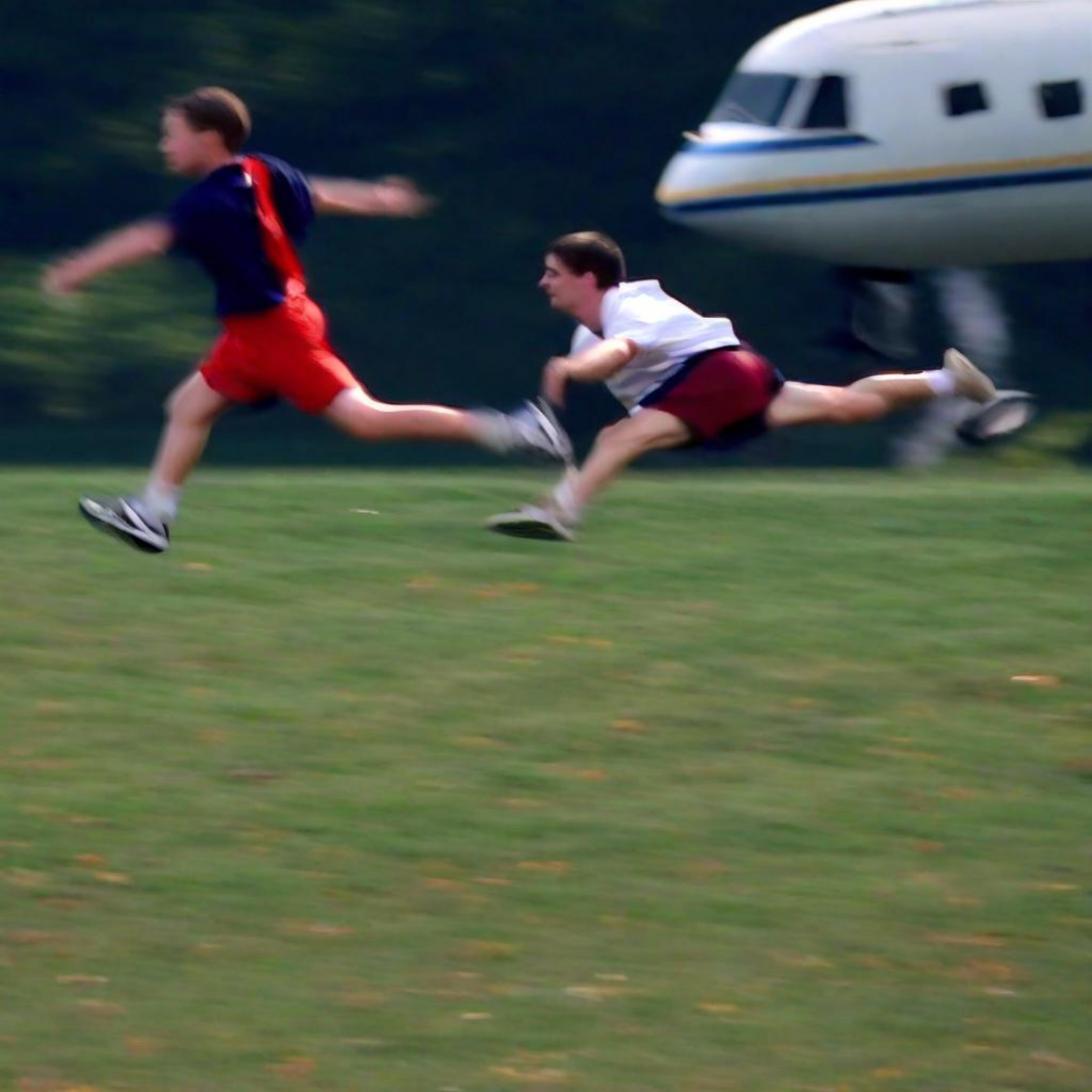}\\
\midrule
Emu2-G &
\centering \includegraphics[align=c,width=0.18\textwidth]{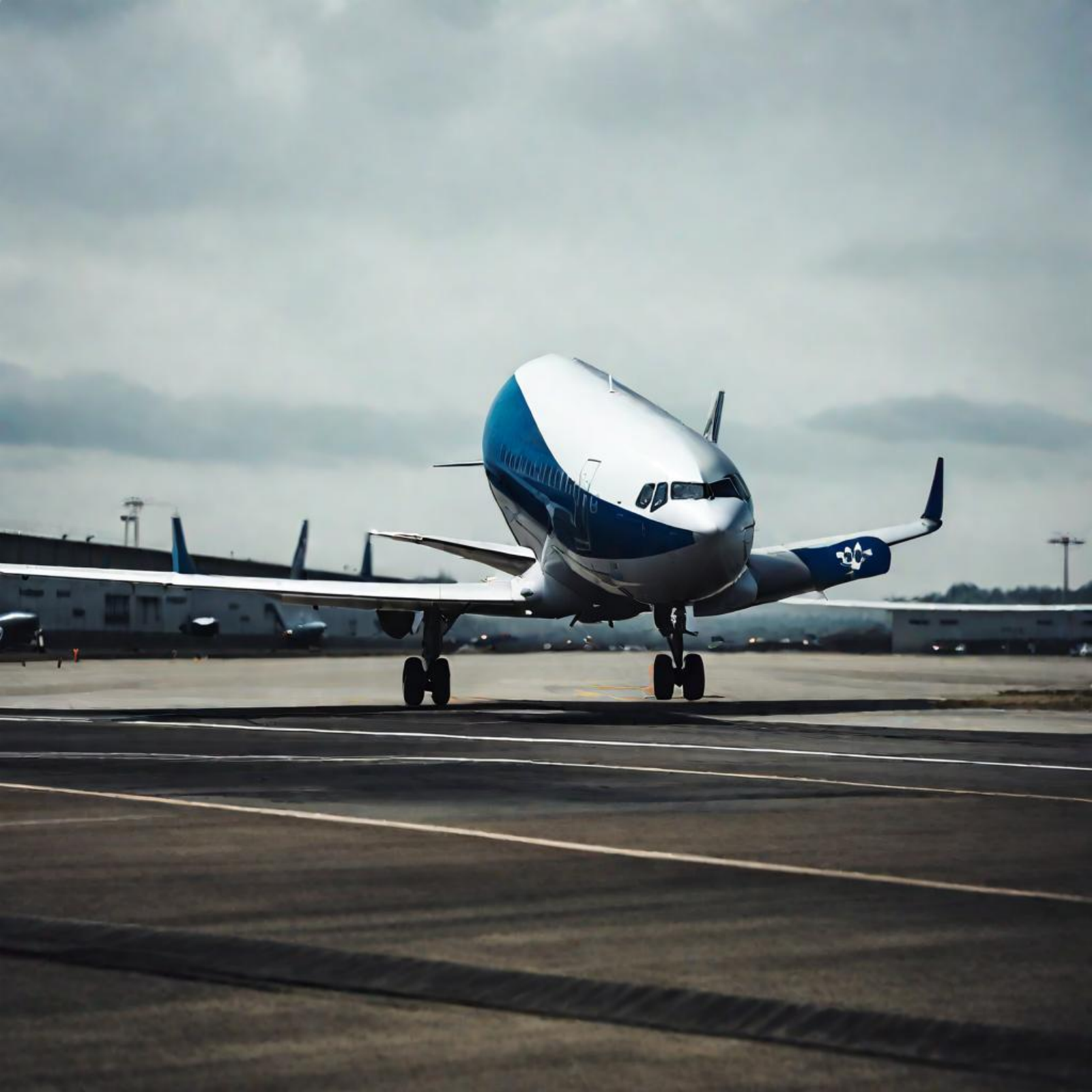}&
\centering \includegraphics[align=c,width=0.18\textwidth]{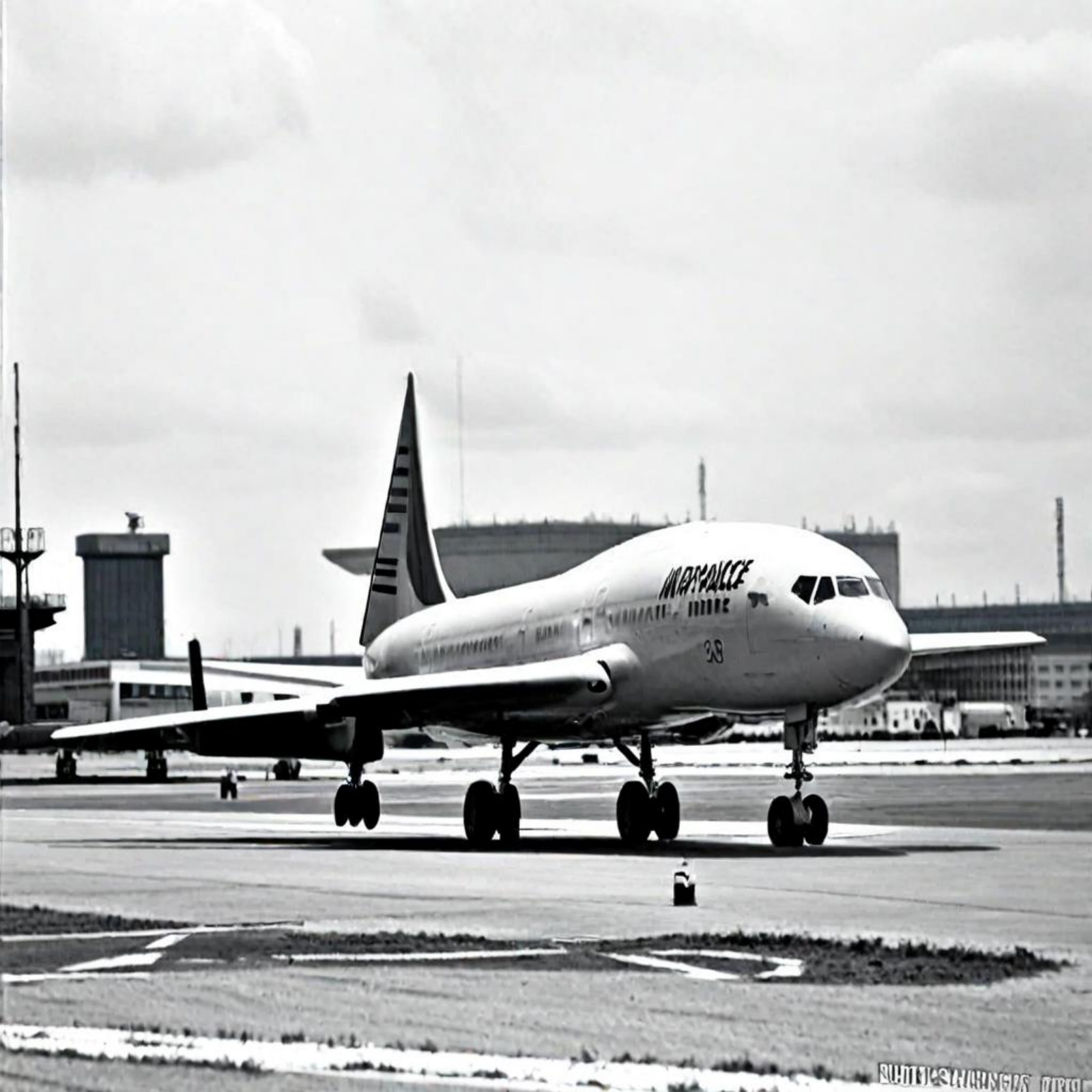}&
\centering \includegraphics[align=c,width=0.18\textwidth]{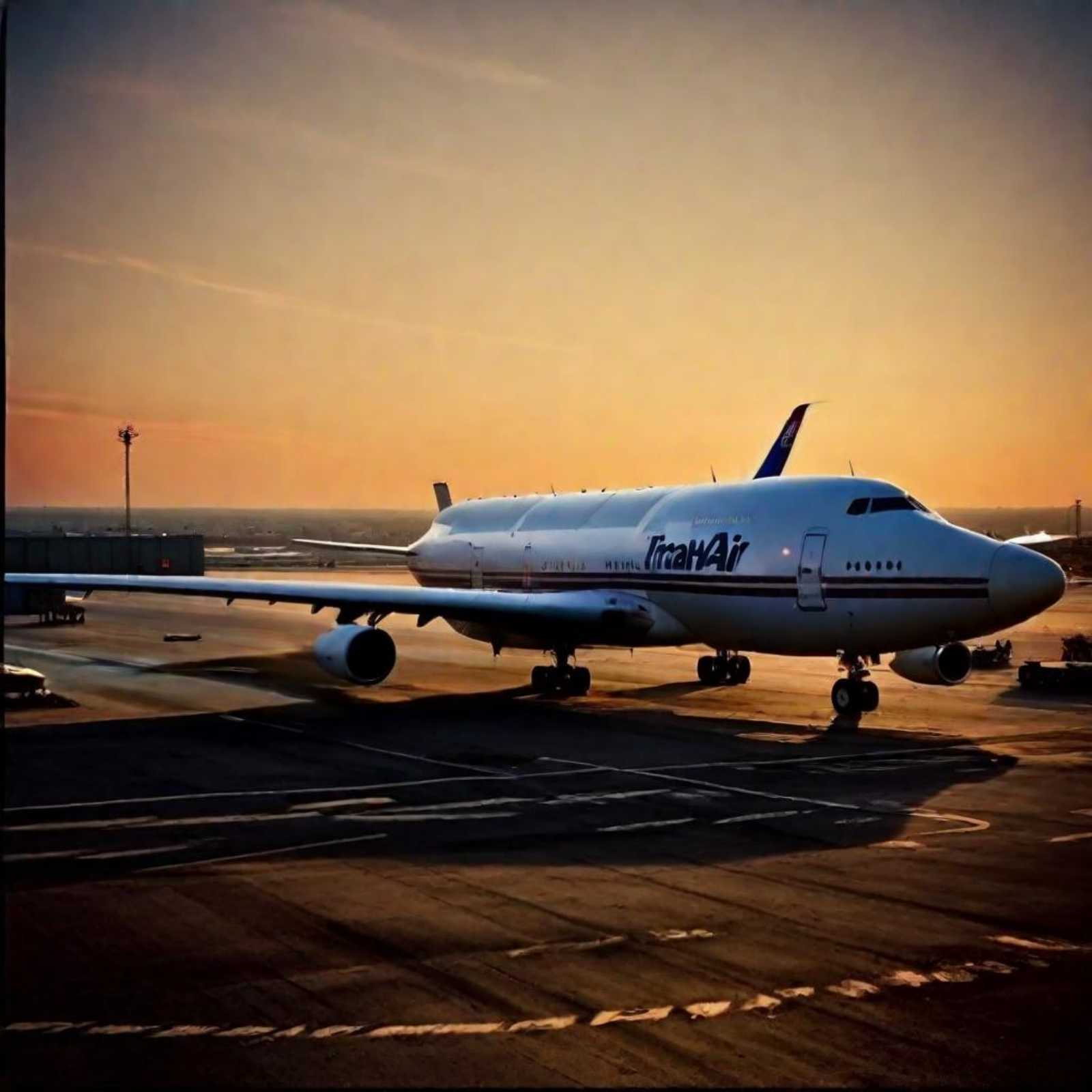}&
\centering\arraybackslash \includegraphics[align=c,width=0.18\textwidth]{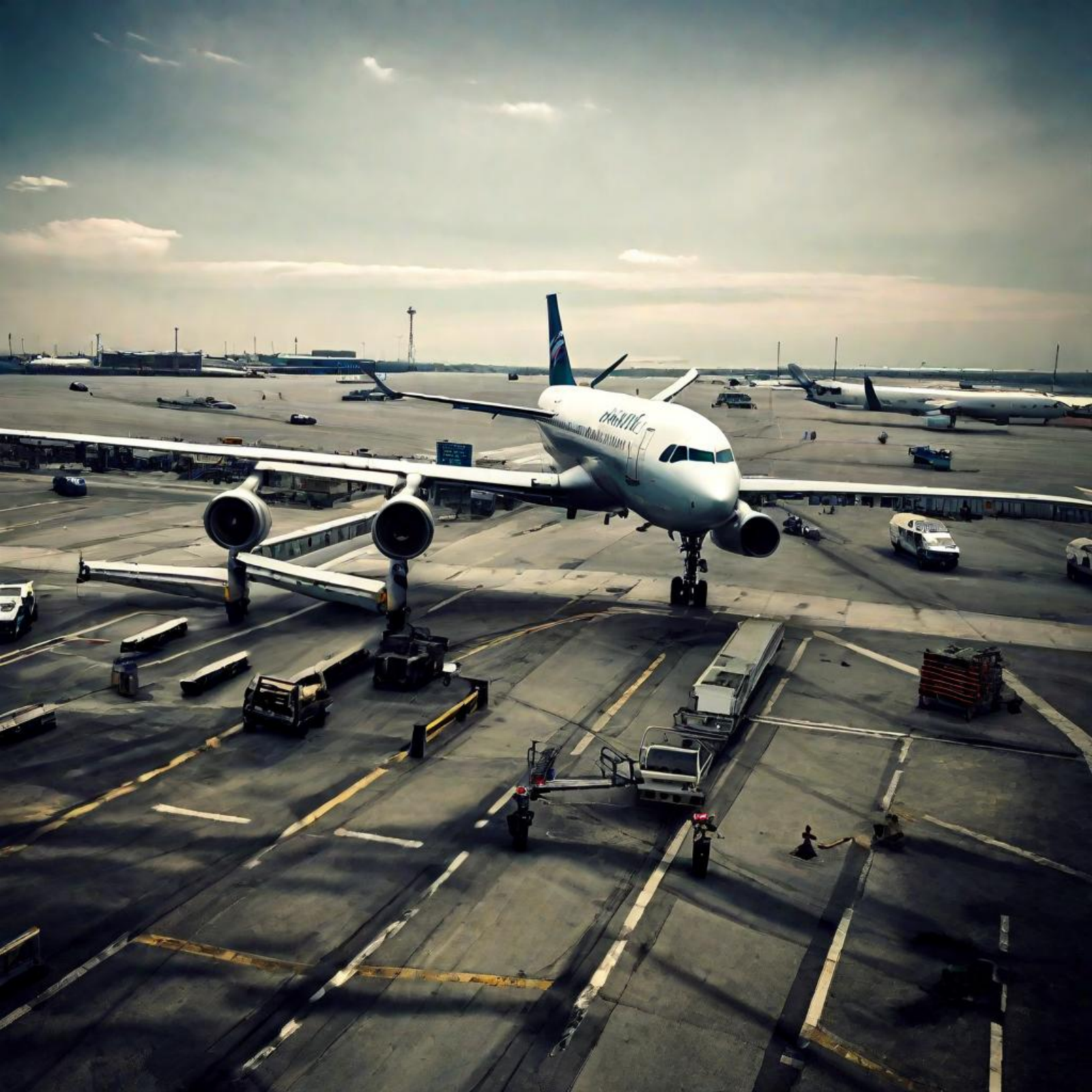}\\
\bottomrule
\end{tabular}%
}
\vspace{-0.1cm}
\caption{Sample image generation with LaVIT, and Emu2-Gen models in zero-shot, and one-shot settings. The ``Prompt'' row shows  the zero-shot caption query as well as retrieved image-caption pairs from CLIP-SF, BLIP-FF and random selection that are included in the prompt as in-context examples.}
\label{tab:image-retriever}
\end{table*}

\subsection{Datasets}
We evaluate UniRAG on image-to-text and text-to-image tasks using the MSCOCO and Fashion200k datasets from the M-BEIR~\cite{uniIR} benchmark. 
 To ensure broad retrieval, we use M-BEIR’s global candidate pool of over 5.5 million entries, including images and texts from all 10 datasets.
This approach allows us to evaluate in a more realistic scenario with a diverse and extensive external database, rather than limiting retrieval to each dataset’s local corpus.

The MSCOCO test set includes about 25k captions for 5k unique images of common objects. For caption generation, we use all 5k images as queries.
However, due to the lengthy image generation time (over 12 seconds per image), we randomly sample one caption query for each image.
To maintain consistency, all image generation runs use this same sample set.
We analyze the effects of this sampling in detail in~\Cref{sec:sampling-effect}.
To demonstrate UniRAG's effectiveness in domain-specific tasks, we also evaluate it on the test set of Fashion200k dataset from M-BEIR.
This set includes about 1.7k caption and 4.9k image queries, all from the fashion domain.

\subsection{Configuration Details}

\begin{table}[thb]
\centering
\resizebox{\columnwidth}{!}{%
\begin{tabular}{lc|ccccccc}
\toprule
\toprule
           & 
\multicolumn{1}{l}{
\multirow{2}{*}{k}} & 
\multicolumn{4}{c} {
\multirow{2}{*}
{\begin{tabular}
[c]{@{}c@{}c@{}c@{}} BLEU- \\
\hspace{-0.1cm}1 \hspace{0.8cm} 2 \hspace{0.7cm} 3 \hspace{0.7cm} 4 \\
\end{tabular}}
} &
\multicolumn{1}{l}{
\multirow{2}{*}{CIDEr}} & 
\multicolumn{1}{l}{
\multirow{2}{*}{ROUGE}} &
\multicolumn{1}{l}{
\multirow{2}{*}{SPICE}}  \\
           & 
\multicolumn{1}{l}{} & 
\multicolumn{4}{c}{} &
\multicolumn{1}{l}{} & 
\multicolumn{1}{l}{} &
\multicolumn{1}{l}{}  \\
\midrule
\midrule
(1a) CLIP-SF & 1 & 88.5 & 84.2 & 81.5 & 79.9 & 215.3 & 83.2 & 37.6\\
\cmidrule{2-9}
(1b) BLIP-FF & 1 & 89.4 & 84.6 & 81.5 & 79.5 & 218.2 & 83.9 & 37.6\\
\midrule
(2) LLaVA & 0 & 58.0 & 39.9 & 27.0 & 18.0 & 64.3 & 42.2 & 17.1\\
\cmidrule{2-9}
(2a) CLIP-SF +& 1 & 80.1 & 70.7 & 63.9 & 59.0 & 162.9 & 70.3 & 31.1\\
\hspace{12pt} LLaVA& 5 & 75.1 & 61.7 & 50.7 & 42.4 & 128.9 & 60.6 & 24.8\\
\cmidrule{2-9}
(2b) BLIP-FF + & 1 & \textbf{81.0} & \textbf{71.5} & \textbf{64.5} & \textbf{59.4} & \textbf{166.9} & \textbf{71.1} & \textbf{31.1}\\
\hspace{12pt} LLaVA& 5 & 75.5 & 61.9 & 50.7 & 42.2 & 129.0 & 60.6 & 25.0\\
\cmidrule{2-9}
(2c) Random + & 1  & 62.6 & 45.7 & 32.7 & 23.1 & 79.2 & 47.5 & 16.9\\
\hspace{12pt} LLaVA & 5 & 46.1 & 27.6 & 16.1 & 9.9 & 27.9 & 36.1 & 7.0\\
\midrule
(3) Gemini-P & 0 & 45.1 & 29.5 & 19.4 & 12.8 & 43.8 & 30.4 & 11.8\\
\cmidrule{2-9}
(3a) CLIP-SF + & 1 & 64.6 & 47.4 & 34.9 & 26.1 & 83.3 & 47.9 & 20.2\\
\hspace{12pt} Gemini-P & 5 & 69.4 & 52.4 & 38.9 & 28.6 & \textbf{96.4} & \textbf{51.6} & \textbf{21.7}\\
\cmidrule{2-9}
(3b) BLIP-FF + & 1 & 64.3 & 47.1 & 34.4 & 25.4 & 83.4 & 48.2 & 20.2\\
\hspace{12pt} Gemini-P & 5 & \textbf{69.9} & \textbf{53.1} & \textbf{39.6} & \textbf{29.4} & 94.4 & 49.9 & 21.0\\
\cmidrule{2-9}
(3c) Random + & 1 & 64.0 & 45.9 & 32.1 & 22.2 & 79.4 & 46.6 & 18.3\\
\hspace{12pt} Gemini-P & 5 & 69.0 & 51.1 & 36.8 & 25.9 & 92.4 & 50.7 & 20.5\\
\midrule
(4) GPT-4o & 0 & 59.6 & 40.1 & 26.4 & 17.0 & 68.3 & 44.9 & 18.5\\
\cmidrule{2-9}
(4a) CLIP-SF + & 1 & 56.6 & 39.0 & 26.8 & 18.7 & 63.3 & 45.5 & 17.7\\
\hspace{12pt} GPT-4o & 5 & 65.4 & 48.4 & 35.2 & .7 & 87.5 & 50.3 & 20.5\\
\cmidrule{2-9}
(4b) BLIP-FF + & 1 & 57.9 & 40.3 & 28.0 & 19.8 & 67.1 & 46.6 & 18.7\\
\hspace{12pt} GPT-4o & 5 & \textbf{66.4} & \textbf{49.2} & \textbf{36.1} & \textbf{26.3} & \textbf{90.4} & \textbf{51.0} & \textbf{21.1}\\
\cmidrule{2-9}
(4c) Random + & 1 & 60.2 & 42.0 & 28.6 & 19.2 & 72.5 & 46.5 & 18.7\\
\hspace{12pt} GPT-4o & 5 & 55.8 & 38.2 & 26.1 & 17.9 & 61.1 & 42.2 & 14.5\\
\bottomrule
\end{tabular}%
}
\caption{Caption generation with LLaVA, Gemini-Pro and GPT-4o models on the MSCOCO dataset using UniIR's CLIP-SF and BLIP-FF as retrievers.
Column $k \in \{0, 1, 5\}$ shows the number of few-shot examples in each run.
}
\label{tab:image-captioning}
\end{table}

We use the LLaVA, Gemini-Pro, and GPT-4o models for generating captions, and LaVIT and Emu2-Gen for image generation.
For both tasks, we experiment with adding $k \in \{0, 1, 5\}$ retrieved image-caption pairs as in-context examples, where $k=0$ establishes the baseline effectiveness of generator models without random or RAG in-context examples.
Since LLaVA does not allow multiple images as input, we vertically merge all few-shot candidate images (and the query image for caption generation) into a single image.
The final merged image's width matches the widest image, while its height is the total of all individual heights.
\Cref{fig:llaval-merged-image} shows a sample merged image with $k=1$ examples.
Each image is then labeled as $[n]$ in the prompt, where $n$ indicates its position in the final merged image.
For all other generator models, the in-context examples are provided in the following interleaved format: $\{\textrm{image}_1, \textrm{text}_1, \ldots ,\textrm{image}_k,  \textrm{text}_k\}$.

We use LLaVA’s default configuration\footnote{\url{https://github.com/huggingface/transformers/blob/main/src/transformers/models/llava/configuration_llava.py}} for all its inferences, with a batch size of four for zero-shot and two for few-shot generation.
Zero-shot caption generation on a single NVIDIA RTX A6000 GPU averages around 5 seconds, while few-shot takes 2-3 times longer.
For Gemini-Pro and GPT-4o, we utilize Vertex AI’s \textit{generate content} and OpenAI’s \textit{chat completion} APIs, respectively, setting their $\texttt{max\_new\_token}$ parameter to 400. More details on their cost estimates can be found in Appendix~\ref{app:cost-estimate}.
For LaVIT, we follow the sample configuration provided,\footnote{\url{https://github.com/jy0205/LaVIT/blob/main/LaVIT/text2image_synthesis.ipynb}} generating a 1024×1024 image in about 12 seconds on a single NVIDIA RTX A6000 GPU.
Similarly, we use Emu2-Gen's default settings.\footnote{\url{https://github.com/baaivision/Emu/tree/main/Emu2\#emu2-gen}}
Since it has 37 billion parameters, we run its ``bf16'' variant on an NVIDIA H100 SXM from the Lambda GPU provider.
Image generation with this setting takes about 8 seconds for zero-shot and one-shot, and 13 seconds for five-shot.
We encountered CUDA out-of-memory errors for 36 out of 5k captions during five-shot image generation with Emu2-Gen, so we limited the prompt to the first four pairs in those cases.

For each task, we report the most commonly used metrics:
For caption generation, we include $n$-gram based metrics, BLEU(1-4)~\cite{papineni-etal-2002-bleu}, CIDEr~\cite{vedantam2015cider}, and ROUGE~\cite{lin-2004-rouge}, as well as the scene-graph-based metric, SPICE.
For image generation, we report FID to measure image fidelity, CLIP Score~\cite{clipscore} for alignment with the caption prompt, and Inspection Score (IS)~\cite{inceptionscore} for overall quality. Appendix~\ref{sec:metrics} explains these metrics in more detail.

\section{Evaluation Results and Analysis}
\label{sec:results}
\begin{table}[b]
\centering
\resizebox{\columnwidth}{!}{%
\begin{tabular}{lc|ccccccc}
\toprule
\toprule
           & 
\multicolumn{1}{l}{
\multirow{2}{*}{k}} & 
\multicolumn{4}{c} {
\multirow{2}{*}
{\begin{tabular}
[c]{@{}c@{}c@{}c@{}} BLEU- \\
\hspace{-0.1cm}1 \hspace{0.8cm} 2 \hspace{0.7cm} 3 \hspace{0.7cm} 4 \\
\end{tabular}}
} &
\multicolumn{1}{l}{
\multirow{2}{*}{CIDEr}} & 
\multicolumn{1}{l}{
\multirow{2}{*}{ROUGE}} &
\multicolumn{1}{l}{
\multirow{2}{*}{SPICE}}  \\
           & 
\multicolumn{1}{l}{} & 
\multicolumn{4}{c}{} &
\multicolumn{1}{l}{} & 
\multicolumn{1}{l}{} &
\multicolumn{1}{l}{}  \\
\midrule
\midrule
(1a) CLIP-SF & 1 & 35.9 & 18.8 & 13.2 & 10.8 &  90.2 & 36.6 & 21.7 \\
\cmidrule{2-9}
(1b) BLIP-FF & 1 & 41.8 & 25.3 & 19.2 & 16.6 & 130.4 & 41.7 & 26.5 \\
\midrule
(2) LLaVA    & 0 & 9.7 & 2.5 & 0.5 & 0.0 &   6.7 & 14.0 & 10.0 \\
\cmidrule{2-9}
(2a) CLIP-SF +& 1 & 32.9 & 16.7 & 11.4 & 9.0 & 75.1 & 33.6 & 19.9 \\
\hspace{12pt} LLaVA& 5 & 34.6 & 15.8 & 9.3 & 6.5 & 63.8 & 35.2 & 20.2 \\
\cmidrule{2-9}
(2b) BLIP-FF + & 1 & 38.9 & \textbf{22.8} & \textbf{16.7} & \textbf{14.0} & \textbf{111.1} & .5 & \textbf{24.1} \\
\hspace{12pt} LLaVA& 5 & \textbf{39.2} & 20.3 & 13.2 & 10.0 & 86.7 & \textbf{39.2} & 23.1 \\
\cmidrule{2-9}
(2c) Random + & 1  & 22.8 & 6.3 & 1.6 & 0.6 & 22.3 & 23.4 & 12.2 \\
\hspace{12pt} LLaVA & 5 & 11.9 & 2.6 & 0.8 & 0.3 & 8.0 &  12.6 & 4.8 \\
\bottomrule
\end{tabular}%
}
\caption{Caption generation with LLaVA on the Fashion200k dataset using UniIR's CLIP-SF and BLIP-FF as retrievers.
Column $k \in \{0, 1, 5\}$ represents the number of few-shot examples for each experiment.}
\label{tab:image-captioning-fashion200k}
\end{table}
\Cref{tab:image-captioning,tab:image-captioning-fashion200k,tab:image-generation,tab:image-generation-fashion200k} present results for caption and image generation tasks on MSCOCO and Fashion200k datasets.
Rows 1* from each table measure the effectiveness of CLIP-SF and BLIP-FF retrievers. 
They reflect how well the generator models perform if they return the image or caption from the first retrieved example as is. 
To ensure an equal number of images and captions for evaluation, we report retrievers' effectiveness using only their top retrieved image-text pair ($k=1$) for each query.

CLIP-SF and BLIP-FF confuse candidate modalities for up to 7\% and 25\% of MSCOCO queries, respectively, meaning they may retrieve an image for an image-to-text or a text for a text-to-image task.
In our evaluation, when a retrieved candidate has the wrong modality, we use its complement with the correct modality.
Unlike the original UniIR results, which indicated CLIP-SF was more effective for MSCOCO retrieval, this adjustment allows both retrievers to perform similarly on this dataset. For Fashion200k, our findings confirm UniIR's results, showing BLIP-FF is more effective.

The remaining rows in each table compare zero-shot ($k=0$) and few-shot ($k>0$) results of generator models using CLIP-SF (*a), BLIP-FF (*b), and random selection (*c) as retrievers for few-shot generations. The behavior of CLIP-SF and BLIP-FF retrievers in RAG applications matches their standalone effectiveness (comparing rows *a and *b in each table). On MSCOCO, both retrievers continue to perform similarly, while BLIP-FF consistently outperforms CLIP-SF on Fashion200k.

\subsection{Caption Generation}
\Cref{tab:image-captioning} shows the caption generation results for the LLaVA, Gemini-Pro and GPT-4o generator models on the MSCOCO dataset. 
While several metrics are reported in this table as a reference, we focus on SPICE since all other metrics are primarily sensitive to $n$-gram overlaps~\cite{anderson2016spice}.

\begin{table}[thb]
\centering
\resizebox{\columnwidth}{!}{%
\begin{tabular}{lc|ccc}
\toprule
\toprule
           & k & FID $\downarrow$  & CLIP Score $\uparrow$   & IS  $\uparrow$ (SD)                      \\
\midrule
\midrule
(1a) CLIP-SF & 1 & 5.88 & 29.99 & 22.19 (1.29)\\
\cmidrule{2-5}
(1b) BLIP-FF & 1 & 7.01 & 29.97 & 22.77 (1.02) \\
\midrule
(2) LaVIT
& 0 & 64.80 & 26.27 & 16.38 (1.27) \\
\cmidrule{2-5}
(2a) CLIP-SF + & 1 & 23.39 & 30.42 & 24.11 (1.09) \\
\hspace{12pt}LaVIT & 5 & 24.14 & 31.09 & 24.52 (1.08) \\
\cmidrule{2-5}
(2b) BLIP-FF + & 1 & \textbf{23.23} & 30.44 & 24.75 (1.25)\\
\hspace{12pt}LaVIT & 5 & 23.95 & 31.07 & \textbf{25.23 (1.10)}\\
\cmidrule{2-5}
(2c) Random + & 1  & 26.90 & 22.43 & 19.31 (0.90) \\
\hspace{12pt}LaVIT & 5 & 26.46 & \textbf{31.13} & 22.89 (1.35) \\
\midrule
(3) Emu2-G
& 0 & 41.66 & 29.54 & 21.43 (0.73)\\
\cmidrule{2-5}
(3a) CLIP-SF + & 1 & \textbf{30.99} & 30.02 & \textbf{23.51 (0.98)}\\
\hspace{12pt}Emu2-G & 5  & 32.51 & 29.94 & 22.01 (1.02)\\
\cmidrule{2-5}
(3b) BLIP-FF +& 1 & 31.37 & \textbf{30.04} & 23.27 (0.75) \\
\hspace{12pt}Emu2-G & 5 & 33.17 & 29.96 & 22.63 (1.84) \\
\cmidrule{2-5}
(3c) Random + & 1 & 38.49 & 23.62 & 16.86 (1.03)\\
\hspace{12pt}Emu2-G & 5 & 53.92 & 23.74 & 11.51 (0.32)\\

\bottomrule
\end{tabular}%
}
\caption{Image generation with LaVIT  and Emu2-G on the MSCOCO dataset using UniIR's CLIP-SF and BLIP-FF as retrievers.
Column $k \in \{0, 1, 5\}$  shows the number of few-shot examples in each run. For Inception Score (IS), its Standard Deviation (SD) is in parenthesis.
}

\label{tab:image-generation}
\end{table}

As reported in~\cref{tab:image-captioning}, augmenting relevant image-caption pairs ($k>0$) retrieved from the previous stage, boosts the effectiveness of LVLMs regardless of their baseline effectiveness, the number of in-context examples and the choice of retriever model.
The baseline effectiveness ($k = 0$) of LLaVA exceeds that of Gemini-Pro and GPT-4o models (comparing rows 2-4).
Even though all models benefit from RA, this gap further widens when only a single retrieved example ($k=1$) is added to the prompts, making one-shot BLIP-FF + LLaVA the most effective setting and one-shot CLIP-SF + LLaVA a very close runner up.
Further increasing the number of retrieved examples from one to five helps the Gemini-P and GPT-4o models.
However, LLaVA with five retrieved examples is less effective than LLaVA with a single example.

For both LLaVA and GPT-4, adding a single randomly selected example to the prompt slightly improves the generation quality compared to zero-shot generation.
However, increasing the number of randomly selected examples to five degrades the generators' effectiveness to below their zero-shot baselines.
In contrast, for Gemini-Pro, random few-shot examples significantly enhance the quality of generated captions, although they are still less effective than few-shot retrieved examples.
These trends show that UniRAG's model-agnostic effectiveness comes from only including the most relevant examples in-context.

\begin{table}[thb]
\centering
\resizebox{\columnwidth}{!}{%
\begin{tabular}{lc|ccc}
\toprule
\toprule
           & k & FID $\downarrow$  & CLIP Score $\uparrow$   & IS  $\uparrow$ (SD)                      \\
\midrule
\midrule
(1a) CLIP-SF & 1 & 13.53 & 26.72 & 4.48 (0.16) \\
\cmidrule{2-5}
(1b) BLIP-FF & 1 & 13.06 & 27.07 & 4.28 (0.21) \\
\midrule
(2) LaVIT & 0 & 117.81 & 21.83 & \textbf{7.74 (0.58)} \\
\cmidrule{2-5}
(2a) CLIP-SF + & 1 & 36.80 & 28.00 & 4.74 (0.24) \\
\hspace{12pt}LaVIT & 5 & \textbf{30.16} & 27.55 &4.44 (0.21) \\
\cmidrule{2-5}
(2b) BLIP-FF + & 1 & 36.68 & \textbf{28.22} & 4.31 (0.16) \\
\hspace{12pt}LaVIT & 5 & 30.24 & 27.50 & 4.36 (0.36) \\
\cmidrule{2-5}
(2c) Random + & 1  & 44.88 & 23.72 & 4.39 (0.27) \\
\hspace{12pt}LaVIT & 5 & 38.89 & 26.15 & 4.13 (0.18) \\

\bottomrule
\end{tabular}%
}
\caption{Image generation with LaVIT on the Fashion200k's datasetg UniIR's CLIP-SF and BLIP-FF as retrievers. 
Column $k \in \{0, 1, 5\}$  represents the number of few-shot examples for each experiment. For Inception Score (IS), its Standard Deviation (SD) is reported in parenthesis.}
\label{tab:image-generation-fashion200k}
\end{table}

\Cref{tab:image-captioning-fashion200k} shows the caption generation results for the Fashion200k dataset. To reduce the cost of API calls, we only report results for the LLaVA model in this experiment.
LLaVA performs poorly in zero-shot scenarios due to its limited knowledge of the domain-specific Fashion200k dataset. 
While randomly selected few-shot examples can help LLaVA grasp the task intent and the required level of detail for describing fashion products, the real improvement occurs when relevant examples are included in the prompts. 
Because many entities in the Fashion200k dataset are quite similar, the generator often repeats the in-context caption of a similar image for its query image, leading to effectiveness that is close to that of the baseline retriever models.
Please see Appendix~\ref{app:fashion200-visual} for sample visualizations.

\subsection{Image Generation}
\Cref{tab:image-generation} shows the image generation results for the LaVIT and Emu2-Gen (denoted as Emu-G in the table) generator models on the MSCOCO dataset. 
As expected, both LaVIT with a 7b backbone LLM and Emu2-Gen with a 33b backbone LLM learn from in-context examples and generate better images when text-image pair examples are included in the prompt.
However, the amount of improvement is different for each model and it varies across different few-shot settings.
While Emu2-G is more effective zero-shot, LaVIT's superior in-context learning ability significantly improves its few-shot generation and makes it the more effective model in all few-shot settings.

\begin{table}[thb]
\centering
\resizebox{0.9\columnwidth}{!}{%
\begin{tabular}{lc|ccc}
\toprule
\toprule
           & k & FID $\downarrow$  & CLIP Score $\uparrow$   & IS  $\uparrow$ (SD)                      \\
\midrule
\midrule
Caption & 0 & 64.80 & 26.27 & 16.38 (1.27)  \\
\hspace{7pt} Set 1 & 1 & 23.39 & 30.42 & 24.11 (1.09)  \\
& 5 & 24.14 & 31.09 & 24.52 (1.08)  \\
\midrule
Caption & 0 & 64.46 & 26.08 & 15.83 (0.86)  \\
\hspace{7pt} Set 2 & 1 & 23.24 & 30.38 & 24.78 (1.97)  \\
& 5 & 24.22 & 31.03 & 24.58  (0.75)  \\
\bottomrule
\end{tabular}
}%

\caption{Image generation with CLIP-SF $+$ LaVIT on two sets of 5k captions sampled from the MSCOCO dataset. Column $k \in \{0, 1, 5\}$ represents the number of few-shot examples for each experiment.
}

\label{tab:sampling-effect}
\end{table}

Both models generate their best results with a single most relevant RAG few-shot example $(k=1)$, regardless of the retriever model. Further increasing the number of RAG examples to $k=5$ has a negligible impact, modestly degrading the image fidelity in favor of slightly higher CLIP and/or Inception scores (IS).
However, the two models behave differently for random few-shot examples; while LaVIT is still mostly effective with randomly selected examples, Emu2-G seems to get confused by too many random examples and its $k=5$ random results are much worse than its zero-shot baseline.

\Cref{tab:image-generation-fashion200k} presents the image generation results for LaVIT on the Fashion200k dataset. 
As expected, FID and CLIP Score show trends similar to those in~\Cref{tab:image-captioning}, with random and RAG examples incrementally enhancing the fidelity of generated images to the ground-truth (lower FID) and improving the alignment of generated images with caption queries (higher CLIP Score). 
However, the Inception Score (IS) is highest in the zero-shot setting. This is due to the low diversity of the Fashion200k dataset; generating more faithful images in this context leads to lower diversity among the generated images, resulting in a lower IS.
Another notable observation is that, with RAG examples, LaVIT achieves a CLIP Score higher than those of the retrievers. This suggests that the generated images align better with the caption queries than the images retrieved from the dataset. Please see Appendix~\ref{app:fashion200-visual} for sample visualizations.

\subsection{Effect of Sampling}
\label{sec:sampling-effect}
For each image in M-BEIR's MSCOCO test set, we have sampled a single caption (Set 1 Captions in~\Cref{tab:sampling-effect}) as a query.
To study the sensitivity to sampling, we repeated the sampling process one more time (Set 2 Captions in~\Cref{tab:sampling-effect}).
The two sample sets have only 1030 captions in common but both of them cover all 5k distinct images in the dataset.
We used CLIP-SF + LaVIT in this ablation study as the retriever-generator pipeline.

\begin{table}[thb]
\centering
\resizebox{\columnwidth}{!}{%
\begin{tabular}{l|ccc}
\toprule
\toprule
        & Baseline  & MSCOCO & Fashion200k \\
\midrule
\midrule
CLIP-SF + & Ground-truth & 23.39 &  36.80 \\
\hspace{12pt}LaVIT & Retrieved & 25.49 &  37.13 \\
\cmidrule{2-4}
BLIP-FF + & Ground-truth & 23.23 & 36.68 \\
\hspace{12pt}LaVIT & Retrieved & 25.17 & 36.34 \\

\bottomrule
\end{tabular}%
}
\caption{Comparison of FID measured in two modes: (a) generated images vs. retrieved images and (b) generated images vs. ground-truth. The images are generated by LaVIT using one-shot examples retrieved by CLIP-SF and BLIP-FF retrievers.
}

\label{tab:fid_vs_retrieved_images}
\end{table}

As shown in~\Cref{tab:sampling-effect}, FID and CLIP Score for all $k$ values are very similar between the two runs with less than one percent difference.
However, the reported Inception Scores and their standard deviations vary slightly more, but still within the confidence interval of one another.
We believe this difference is related to the inherent uncertainty of the metric when applied to small dataset sizes.

\subsection{Datasets Comparison}
As shown in~\Cref{tab:image-captioning-fashion200k,tab:image-generation-fashion200k}, smaller models perform poorly on the Fashion200k dataset in both tasks under zero-shot settings, indicating their limited implicit knowledge of the fashion domain.
However, their strong reliance on in-context examples and significant improvement under few-shot RAG settings demonstrate UniRAG’s effectiveness in bridging this knowledge gap.

In contrast, for the MSCOCO dataset (\Cref{tab:image-captioning,tab:image-generation}), where smaller models have more inherent knowledge of common entities, the zero-shot effectiveness is greater and the impact of few-shot RAG examples is less pronounced.
While few-shot RAG examples provide greater benefits for the Fashion200k dataset, this does not imply that the model merely replicates retrieved images in the absence of pre-trained knowledge. As shown in~\Cref{tab:fid_vs_retrieved_images}, the Fashion200k dataset exhibits a significantly higher FID when comparing generated images to the top-one retrieved image, indicating that the model does not simply copy but instead generates distinct outputs.

\section{Conclusion}
We introduced UniRAG as a model agnostic retrieval augmentation technique for Vision-Language (VL) tasks and evaluated it on image-to-text and text-to-image tasks.
We utilized the LLaVA, Gemini-Pro, and GPT-4o models to generate captions for input images in zero-shot and few-shot (with RAG) settings. 
Similarly, we employed the LaVIT and Emu2-Gen models to generate images from input captions.
In the RAG few-shot setting, we leveraged UniIR's CLIP-SF and BLIP-FF models to retrieve relevant image-text pairs and included them as in-context examples. To further showcase the effectiveness of RAG few-shot examples, we also compared them against randomly selected few-shot examples.

Our experimental results on the MSCOCO dataset from M-BEIR indicated that all models, regardless of their zero-shot baseline effectiveness, improved after being exposed to in-context examples. However, the best results were achieved only when relevant retrieved examples were included in the prompts, rather than random text-image pairs. In fact, for the LLaVA, GPT-4o, and Emu2-Gen models, adding five random examples decreased generation quality below their zero-shot baselines.
In contrast, in the RAG few-shot setting, increasing the number of examples from one to five either improved generation quality (for Gemini-Pro and GPT-4o) or had no effect (for LaVIT and Emu2-Gen). For LLaVA, while adding more RAG in-context examples reduced its effectiveness, it still performed significantly better than both the zero-shot and random few-shot prompts.
Our experiments on five LVLMs confirmed that UniRAG with UniIR retrievers is a model-agnostic effective way to improve the generation quality of pre-trained LVLMs by retrieving relevant few-shot examples at inference time.

To assess UniRAG's effectiveness with uncommon domain-specific entities, we evaluated it on the Fashion200k dataset. 
Our results showed that UniRAG is essential to improve generation quality in scenarios where the generator model has limited implicit knowledge about the entities.

\section{Limitations}
While we have open-sourced UniRAG, its usage is still bound by the licenses and the usage agreements of proprietary GPT-4o and Gemini-Pro models used for inference. Furthermore, one of our open-source models LaVIT, is licensed under the LaVIT Community License which requires specific license requests for commercial use.

The primary goal of UniRAG is to enhance the model’s knowledge of specific subject entities, ensuring that retrieved information directly influences the generated output—a common characteristic of all RAG applications.
While retrieving from a diverse database and incorporating multiple few-shot examples can broaden this influence, this approach may be less suitable for applications prioritizing creativity or originality over generation accuracy.
Additionally, relevance is the only criterion used while retrieving candidate pairs for in-context examples.
Deploying this method in real-world applications requires ranking candidates based on facts, potential harms, biases and other important factors for responsible AI, rather than solely relying on relevance.
The added latency and computation cost of retrieval and inference with larger prompts must also be considered for time- or cost-sensitive applications.

This work uses the English-only MSCOCO and Fashion200k datasets for evaluation and experiments with retriever and generator models that are primarily trained on English corpus. Thus, the effectiveness of our proposed method for non-English low-resource languages remains unexplored.
Moreover, generating images of real people necessitates extra caution and extensive auditing due to significant privacy and security implications.
Finally, further research is required to assess the generalizability of the retriever-guided generation to out-of-domain retrieval.
For instance, employing an entity-centric evaluation dataset, which the retriever has not been trained on, could better demonstrate the advantage of retrieval augmentation for multi-modal inference.

\section*{Acknowledgments}
This research was supported in part by the Natural Sciences and Engineering Research Council (NSERC) of Canada.
We also thank Microsoft and Google for providing access to OpenAI LLMs via Azure and Gemini via VertexAI, respectively. Additionally, we appreciate Cong Wei for answering several questions about UniIR and Xueguang Ma for providing valuable feedback.

\bibliography{main}

\clearpage

\appendixtitleon
\appendixtitletocon
\begin{appendices}

\section{Prompt Templates}
\label{sec:prompt-details}
This section shows the generic zero-shot and model-specific few-shot prompt templates for the caption generation task.

\begin{figure}[!tbh]
\tiny
\justifying
\begin{minted}[fontsize=\tiny, frame=lines, frame=single,linenos=false,breaklines,breaksymbol=,escapeinside=||,bgcolor=LightGray]{text}
You are an intelligent helpful assistant AI that is an expert in generating captions for provided images.
I have provided you an image. Generate the caption for this image based on your understanding of the image.
Only respond with the caption; do not say any other words or explain.
\end{minted}
\vspace*{-0.5cm}
\caption{Zero-shot prompting for caption generation.}
\label{fig:zero-shot-prompt}
\end{figure}
\vspace*{-0.5cm}

\begin{figure}[!tbh]
\tiny
\justifying
\begin{minted}[fontsize=\tiny, frame=lines, frame=single,linenos=false,breaklines,breaksymbol=,escapeinside=||,bgcolor=LightGray]{text}
{{merged images}}

You are an intelligent helpful assistant AI that is an expert in generating captions for provided images.

I have provided you {image_num} images. The following {caption_num} captions correspond to the first {caption_num} images.

{{captions}}

Generate the caption for the last remaining image based on your understanding of the last image and provided image-caption examples.
Only respond with the caption; do not say any other words or explain.

\end{minted}
\vspace*{-0.5cm}
\caption{Few-shot prompting for caption generation with LlaVA.}
\label{fig:few-shot-llava-prompt}
\end{figure}
\vspace*{-0.5cm}
\begin{figure}[!tbh]
\tiny
\justifying
\begin{minted}[fontsize=\tiny, frame=lines, frame=single,linenos=false,breaklines,breaksymbol=,escapeinside=||,bgcolor=LightGray]{text}
I have provided you {num} image(s) and their corresponding caption(s) as example(s) and one last image without a caption.

Generate the caption for the last remaining image based on your understanding of the last image and provided image-caption examples.
Only respond with the caption string; do not say any other words or explain.

{{image-caption-pairs}}

\end{minted}
\vspace*{-0.5cm}
\caption{Few-shot prompting for caption generation with Gemini-Pro.}
\label{fig:few-shot-gemini-prompt}
\end{figure}
\vspace*{-0.5cm}
\begin{figure}[!tbh]
\tiny
\justifying
\begin{minted}[fontsize=\tiny, frame=lines, frame=single,linenos=false,breaklines,breaksymbol=,escapeinside=||,bgcolor=LightGray]{text}
You are an intelligent helpful assistant AI that is an expert in generating captions for provided images.

{{image-caption-pairs}}

I have provided you {num} image(s) and their corresponding caption(s) as example(s) and one last image without a caption.

Generate the caption for the last remaining image based on your understanding of the last image and provided image-caption examples.
Only respond with the caption; do not say any other words or explain.
\end{minted}
\vspace*{-0.5cm}
\caption{Few-shot prompting for caption generation with GPT-4o.}
\label{fig:few-shot-gpt-prompt}
\end{figure}
\vspace*{-0.5cm}
\begin{table}[tbh]
    \centering
    \resizebox{\columnwidth}{!}{%
    \begin{tabular}{c|c|c|c}
        \toprule
        \toprule
         \multirow{2}{*}{Model} & \multicolumn{2}{c|}{Input} & Output \\
        \cmidrule{2-4}
          & Text & Image & Text \\
        \midrule
        \midrule
        GPT-4o & \$0.005  & \$0.015 & \$0.005 \\
        Gemini-Pro & \$0.000125 & \$$0.0025^*$ & \$0.000375 \\
        \bottomrule
    \end{tabular}
    }
    \caption{Model API pricing per 1,000 tokens is listed in USD. ($*$) The cost for a Gemini API call with image input is presented per image.}
    \label{tab:cost}
\end{table}
\section{Proprietary Models Cost}
\label{app:cost-estimate}
We used OpenAI and Vertex APIs for generating captions with GPT-4o and Gemini-Pro models, respectively.
\Cref{tab:cost} illustrates the API pricing information for both models. Overall, for the evaluation of the MSCOCO test set which includes 5k image-text pairs, our average cost estimates per run are approximately 45 USD for the GPT-4o model and 35 USD for the Gemini-Pro model. Image captioning experiments in this paper cost about 560 USD in total (seven runs for each model).

\section{Generation Visualization}
This section shows a sample merged image for few-shot caption generation with LLaVA.

\begin{figure}[!tbh]

\begin{tabular}{c|c}
\subfloat[Random]{\includegraphics[width=0.45\columnwidth]{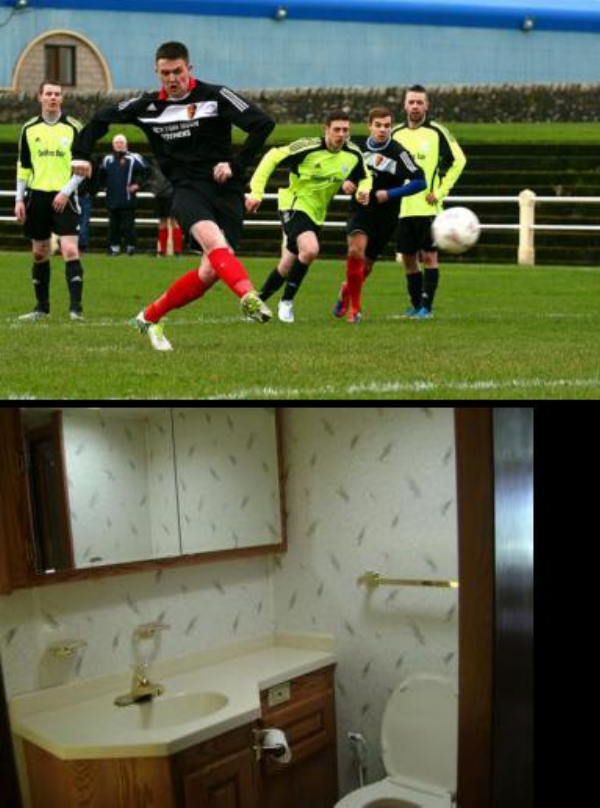}} &
\subfloat[CLIP-SF]{\includegraphics[width=0.45\columnwidth]{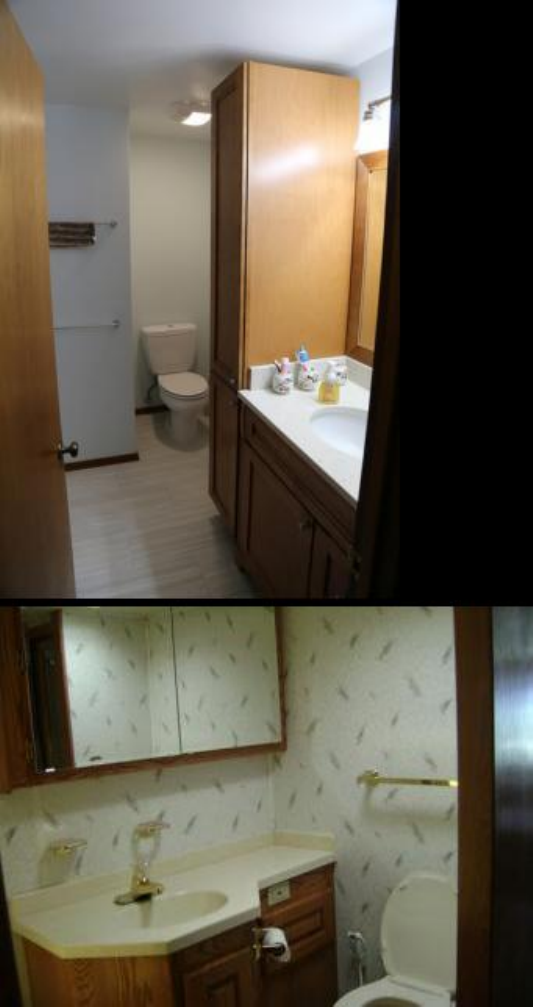}} \\

\end{tabular}%
\caption{A sample image merging for LLaVA prompts. Few-shot images are vertically merged into the query-image for (a) random and (b) CLIP-SF retrievers with $k=1$ examples.}
\label{fig:llaval-merged-image}
\end{figure}

\begin{table*}[thb]
\centering
\resizebox{\textwidth}{!}{%
\begin{tabular}{c | p{5.8cm} | p{5cm} p{5.5cm} p{5.3cm}}
\toprule
&
\centering \textbf{Image Query} &
\multicolumn{3}{c}{\textbf{Top retrieved image-caption pair $(k=1)$}}\\
&
\centering Zero-shot &
\centering CLIP-SF &
\centering BLIP-FF &
\centering\arraybackslash Random\\
\midrule
\multirow{4}{*}{{Prompt}} &
\centering \includegraphics[align=c,width=0.18\textwidth]{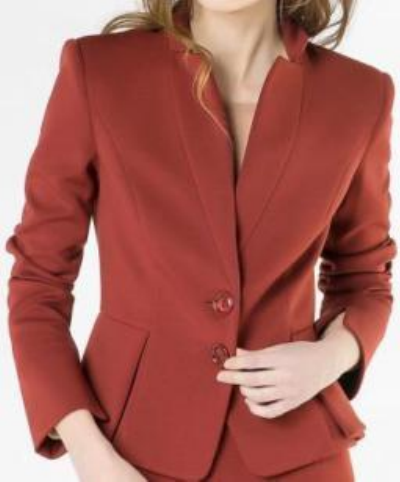}
&
\centering \includegraphics[align=c,width=0.18\textwidth]{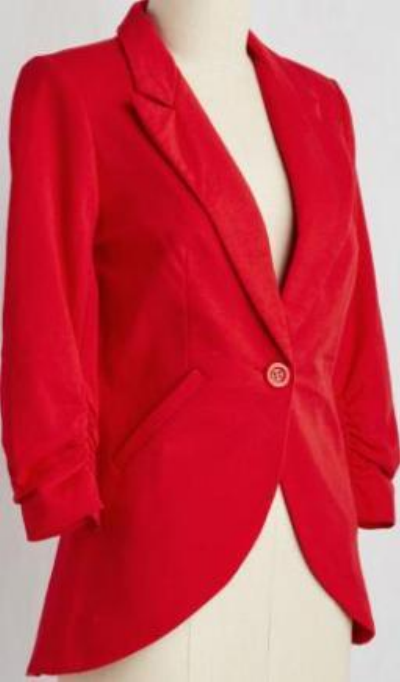}
&
\centering \includegraphics[align=c,width=0.18\textwidth]{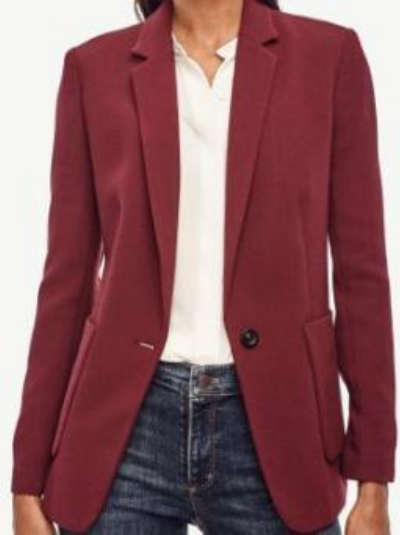}
&
\centering\arraybackslash \includegraphics[align=c,width=0.18\textwidth]{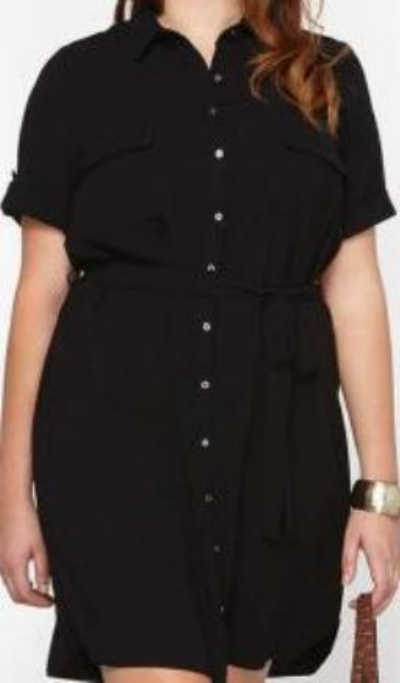}
\\
\cmidrule{2-5}
&
\centering \multirow{3}{*}{{-}}&
Multicolor fine sandy blazer red.&
Red petite refined one button blazer.&
Dp curve black collared shirt dress.\\
\midrule
\midrule
LLaVA &
Effortless style in a vibrant red blazer.&
Red blazer with a white collar.&
Red blazer with a high neckline.&
Red blazer with black collar.\\
\bottomrule
\end{tabular}%
}
\caption{Sample caption generation with LLaVA model on the Fashion200k dataset in zero-shot and one-shot settings. The ``Prompt'' row shows the zero-shot image query as well as retrieved image-caption pairs from CLIP-SF, BLIP-FF and random selection that are included in the prompt as in-context examples.}
\label{tab:caption-retriever-fashion200k}
\end{table*}


\begin{table*}[!thb]
\centering
\resizebox{\textwidth}{!}{%
\begin{tabular}{c | p{5cm} | p{4.3cm} p{4.6cm} p{6cm}}
\toprule
&
\centering \textbf{Caption Query} &
\multicolumn{3}{c}{\textbf{Top retrieved image-caption pair $(k=1)$}}\\
&
\centering Zero-shot &
\centering CLIP-SF &
\centering BLIP-FF &
\centering\arraybackslash Random\\
\midrule
\multirow{4}{*}{{Prompt}} &
\centering -&
\centering \includegraphics[align=c,width=0.18\textwidth]{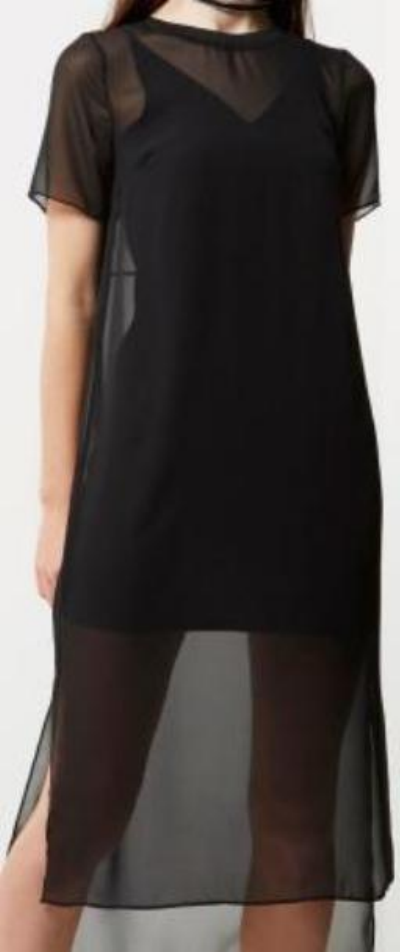}
&
\centering \includegraphics[align=c,width=0.18\textwidth]{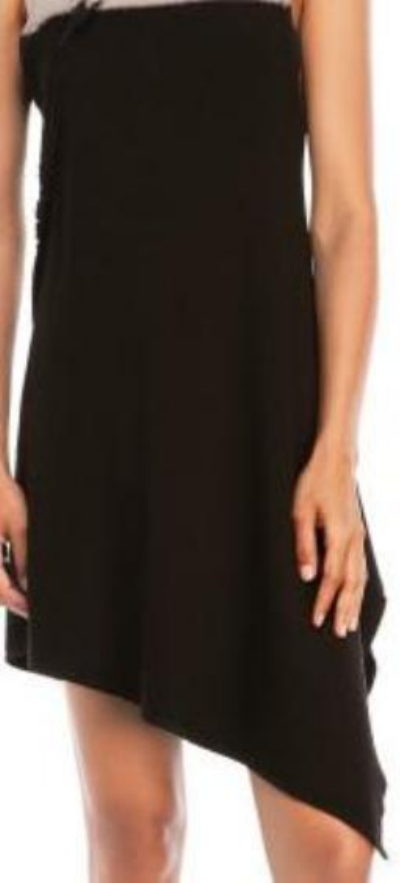}
&
\centering\arraybackslash \includegraphics[align=c,width=0.18\textwidth]{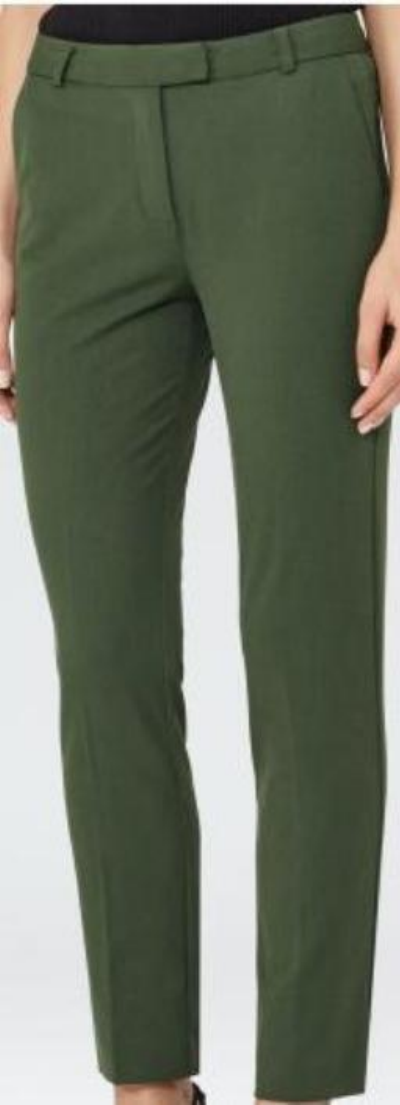}
\\
\cmidrule{2-5}
&
Black asymmetric overlay dress.&
Black sheer panel dress.&
Black asymmetric panel dress.&
Blue joanne cropped tailored trousers.\\
\midrule
\midrule
LaVIT &
\centering \includegraphics[align=c,width=0.18\textwidth]{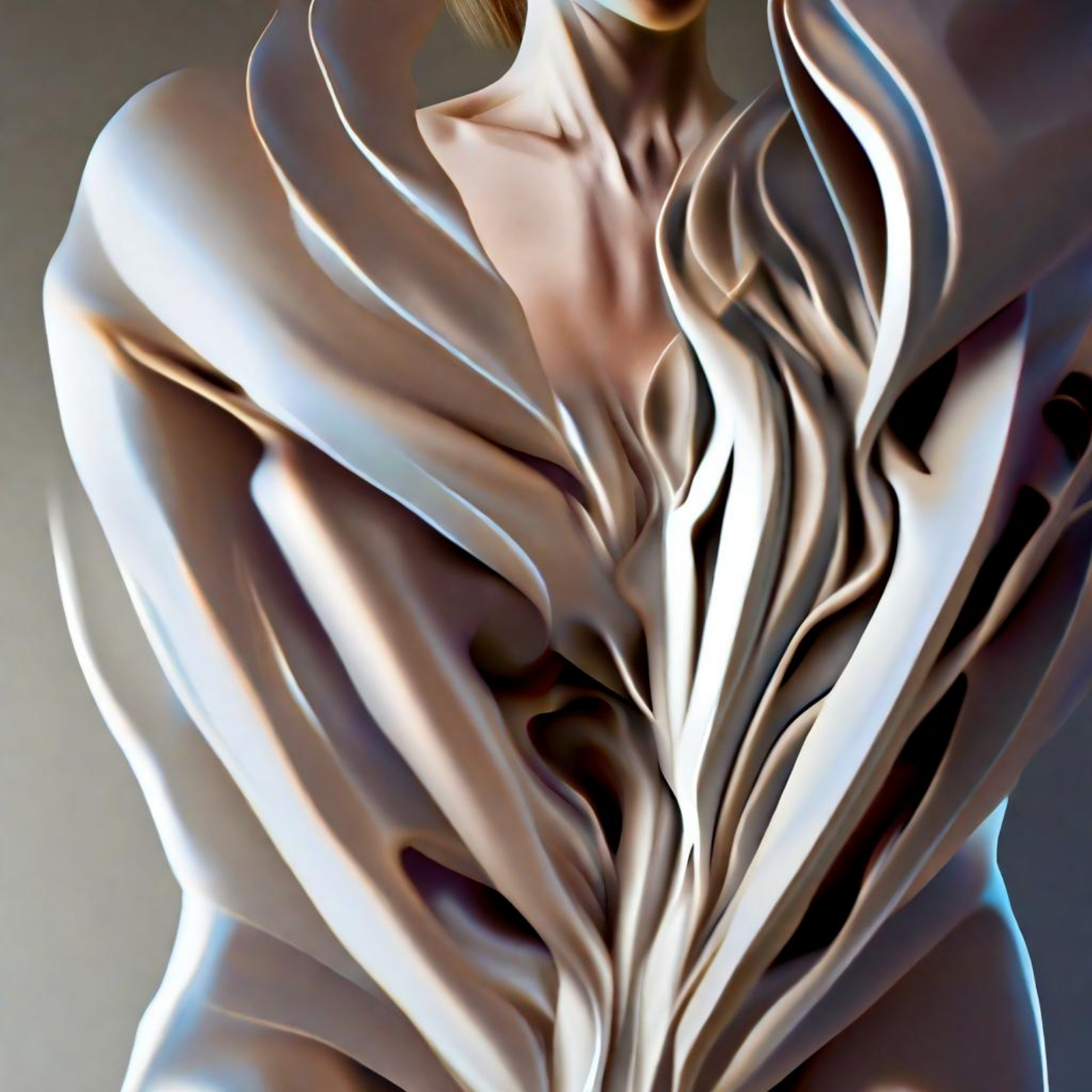}&
\centering \includegraphics[align=c,width=0.18\textwidth]{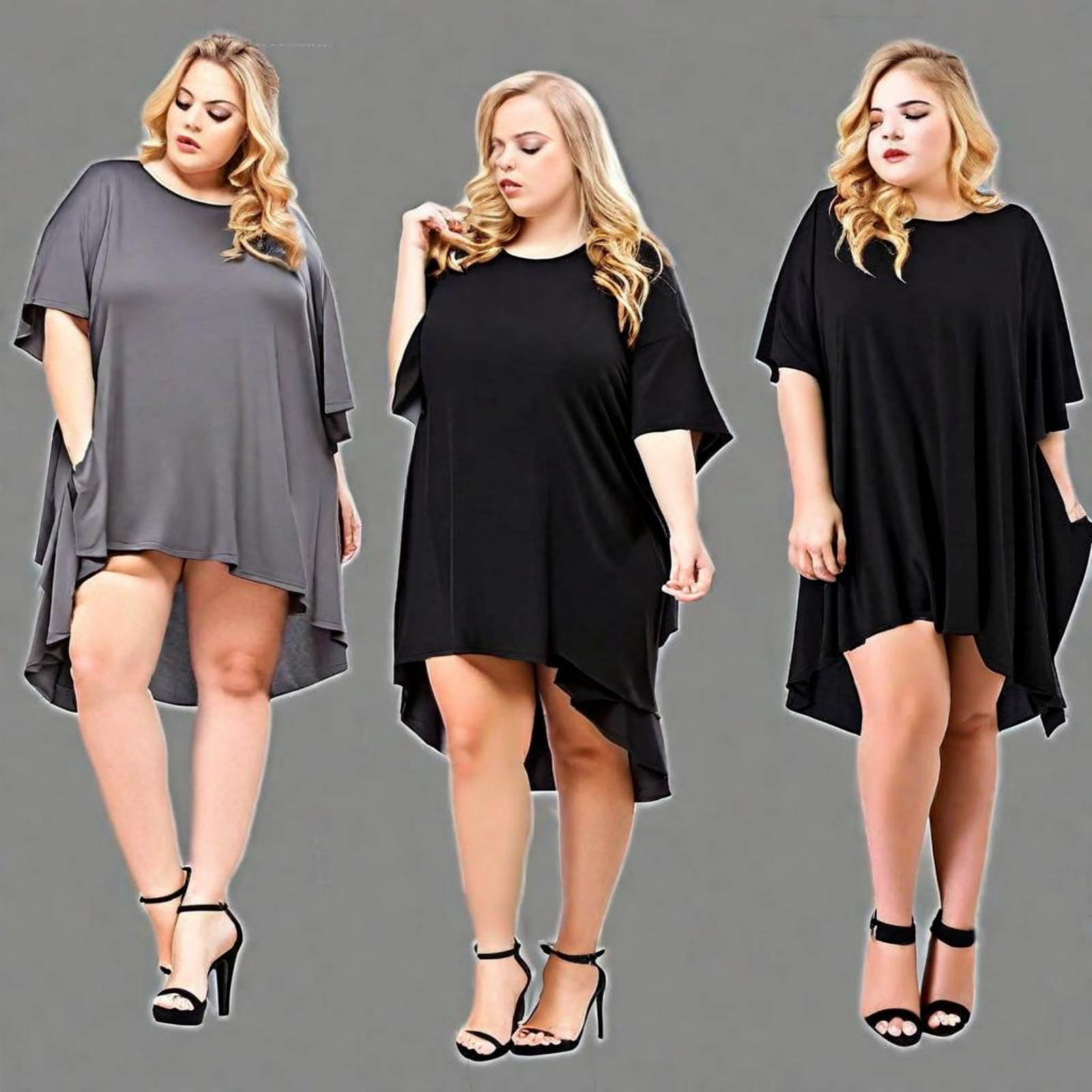}&
\centering \includegraphics[align=c,width=0.18\textwidth]{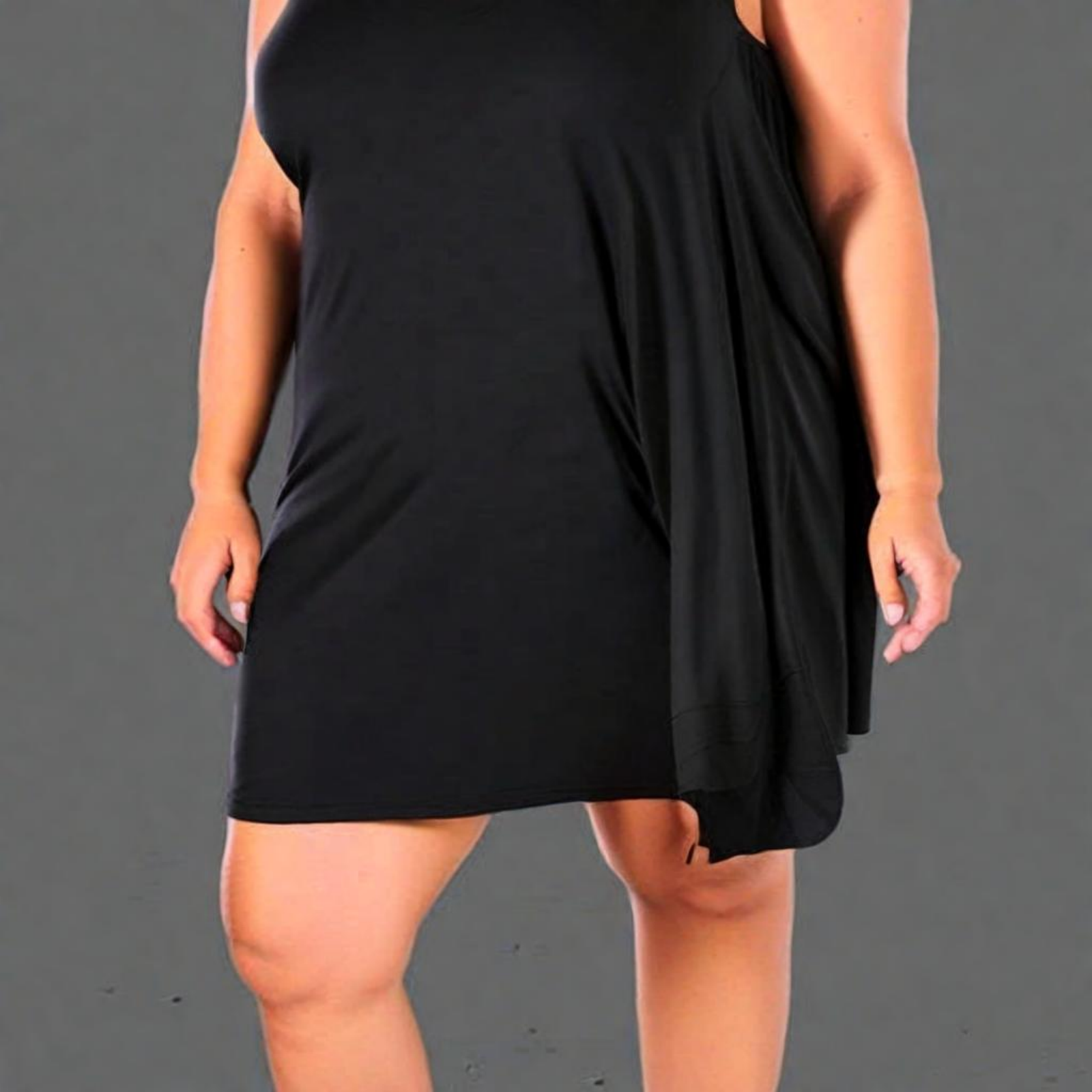}&
\centering\arraybackslash \includegraphics[align=c,width=0.18\textwidth]{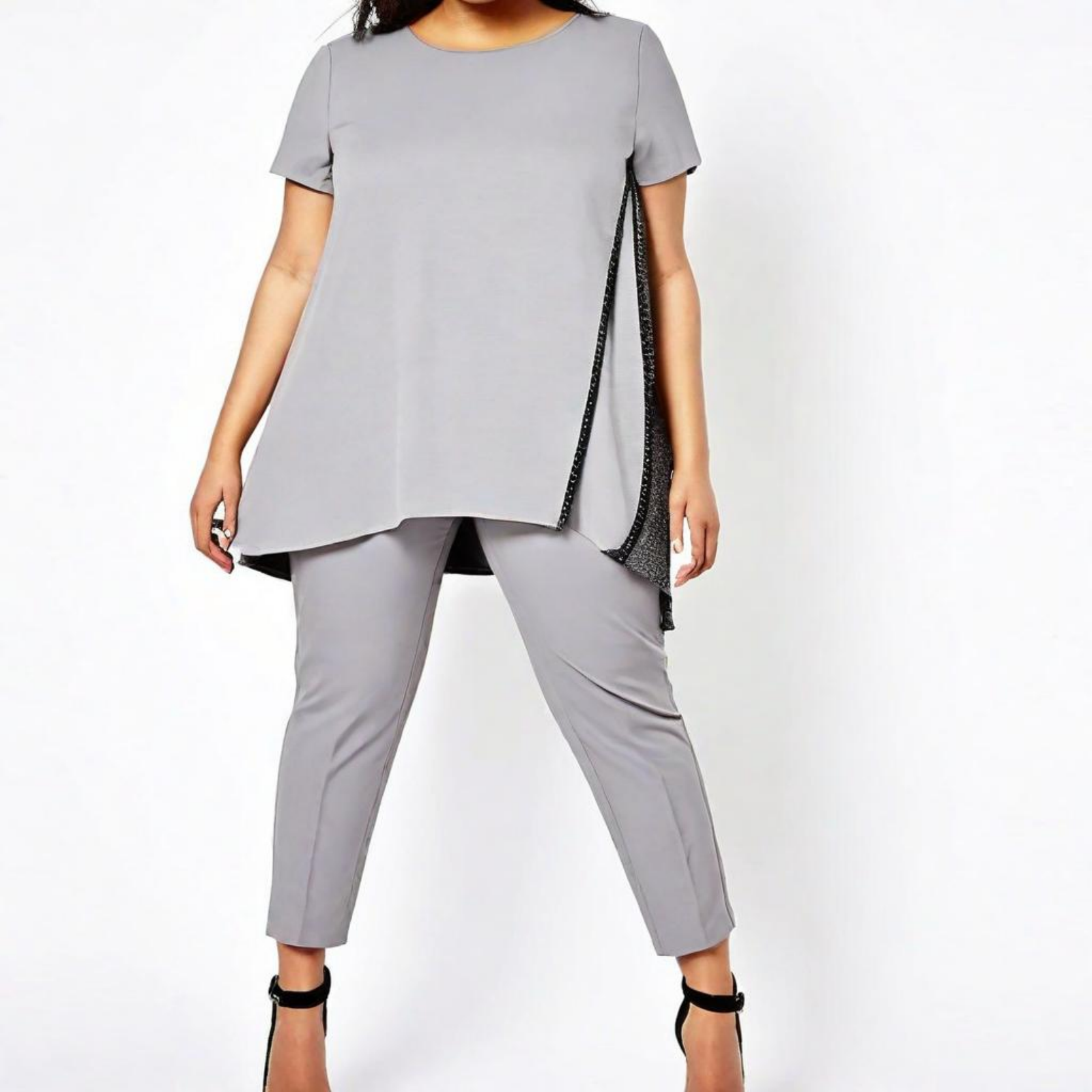}\\
\bottomrule
\end{tabular}%
}
\caption{Sample image generation with LaVIT on the Fashion200k dataset in zero-shot and one-shot settings. The ``Prompt'' row shows  the zero-shot caption query as well as retrieved image-caption pairs from CLIP-SF, BLIP-FF and random selection that are included in the prompt as in-context examples.}
\label{tab:image-retriever-fashion200k}
\end{table*}

\section{Reported Metrics}
\label{sec:metrics}
In our evaluations, we report commonly used metrics for both tasks. For caption generation, we use BLEU (1-4)~\cite{papineni-etal-2002-bleu} and ROUGE~\cite{lin-2004-rouge}, which measure the precision and recall of $n$-grams in the generated captions compared to ground-truth captions. We also employ CIDEr~\cite{vedantam2015cider}, which assesses cosine similarity between $n$-gram vectors from the generated and ground-truth captions, and SPICE~\cite{anderson2016spice}, which compares the semantic content of the generated and ground-truth captions using scene-graph tuples.

For image generation, we use FID~\cite{fid} to measure the KL divergence between the feature vector distributions of generated and ground-truth images; a lower FID indicates greater similarity. Additionally, we calculate the CLIP Score~\cite{clipscore}, which measures cosine similarity between CLIP’s visual and text embeddings for a generated image and its caption query. Lastly, we include the Inception Score (IS)~\cite{inceptionscore} and its Standard Deviation (SD) as another quality indicator for generated images. A higher IS indicates more diversity and confident classification of the generated images. However, IS provides a weaker signal compared to the other two metrics for image generation, as it is less accurate for small datasets.
on{Fashion200k Dataset Visualization}
\label{app:fashion200-visual}
\Cref{tab:caption-retriever-fashion200k,tab:image-retriever-fashion200k} visualize sample caption and image generation for the Fashion200k dataset.

\end{appendices}

\end{document}